\documentclass[aps,prb,reprint,superscriptaddress,longbibliography,nofootinbib,noeprint]{revtex4-2}

\usepackage[T1]{fontenc}
\usepackage{physics}
\usepackage{mathtools,amssymb,slashed,bbm,tensor,verbatim,siunitx,slashed,xfrac}
\usepackage{mathrsfs}
\usepackage[utf8]{inputenc}

\usepackage{tikz}
\usetikzlibrary{decorations.markings,decorations.pathreplacing,arrows,arrows.meta,calc,shapes.arrows,shapes.geometric}
\usepackage{array,multirow}
\usepackage[colorlinks, linkcolor=red,citecolor=blue,urlcolor=blue]{hyperref}
\usepackage[all]{hypcap}

\usepackage[caption=false]{subfig}
\DeclareRobustCommand{\op}[2]{\tensor*{\hat{#1}}{_{#2}}}
\DeclareRobustCommand{\opd}[2]{\tensor*{\hat{#1}}{^{\dagger}_{#2}}}
\DeclareRobustCommand{\d}{\mathrm{d}}

\DeclareRobustCommand{\e}{\mathrm{e}}
\DeclareRobustCommand{\i}{\mathrm{i}}

\DeclareMathOperator{\sign}{\mathrm{sgn}}
\DeclareMathOperator{\diag}{\mathrm{diag}}

\DeclareRobustCommand{\Green}[2]{\,\vcenter{\hbox{\begin{tikzpicture} \newcommand\h{0.21}; \draw (0,-\h)--(-\h,0)--(0,\h); \draw ($(0,-\h)-(0.1,0)$)--($(-\h,0)-(0.1,0)$)--($(0,\h)-(0.1,0)$); \end{tikzpicture}}} #1 ; #2 \vcenter{\hbox{\begin{tikzpicture} \newcommand\h{0.21}; \draw (0,-\h)--(\h,0)--(0,\h); \draw ($(0,-\h)+(0.1,0)$)--($(\h,0)+(0.1,0)$)--($(0,\h)+(0.1,0)$); \end{tikzpicture}}}}

\usepackage{xprintlen}

\begin{document}

\title{
Properties of topological insulators and superconductors under relativistic gravity
}

\author{Patrick J. Wong}

\affiliation{Department of Physics, University of Connecticut, Storrs, Connecticut 06269, USA}

\affiliation{Nordita, Stockholm University and KTH Royal Institute of Technology, Hannes Alfv\'ens v\"ag 12, SE-106~91 Stockholm, Sweden}

\author{Zackary White}

 \affiliation{Department of Physics, University of Connecticut, Storrs, Connecticut 06269, USA}

\author{Alexander V. Balatsky}

\affiliation{Department of Physics, Institute for Materials Science, University of Connecticut, Storrs, Connecticut 06269, USA}

\affiliation{Nordita, Stockholm University and KTH Royal Institute of Technology, Hannes Alfv\'ens v\"ag 12, SE-106~91 Stockholm, Sweden}

\date{\today}

\begin{abstract}
The interplay between the curved spacetimes of general relativity and quantum mechanical systems is an active field of research. However, analysis of relativistic gravitation on extended quantum systems remains understudied. To this end, we study here the effects of a general relativistic curved spacetime on the topological phases of the Su-Schrieffer-Heeger model and Kitaev superconducting wire. We find that the topological states remain robust and well localized. In the topological insulator we find that the energy level of the topological state becomes shifted away from zero according to the gravitational redshift, breaking the system's chiral symmetry. In contrast, the Majorana zero mode of the topological superconductor remains at zero energy. Furthermore, within the topological superconductor, we identify the possibility of a gravitationally induced topological phase transition leading to the formation of a domain wall, shifting one of the boundary Majorana zero modes into the bulk.
\end{abstract}

\maketitle

\section{Introduction}

Gravitation is an interaction which affects all forms of matter. Although a complete quantum theory of gravitation does not yet exist, it has been demonstrated that even the classical description of gravity yields non-trivial effects on quantum systems.

One of the principal effects general relativity imprints onto quantum systems is that of the gravitational redshift, originally predicted for classical light and later experimentally verified for photons, wherein characteristic energy scales become modulated by the strength of the gravitational scalar potential~
\cite{poundrebka,hafelekeating,mueller2010,bothwell2022,zheng2023,afanasiev2024}.
Another imprint of the classical gravitational field on quantum systems is the gravitational Aharonov-Bohm effect. This is again due to gravitational scalar potential and is thus a scalar version of the traditional vector Aharonov-Bohm effect of electromagnetism. This effect too has been measured in recent years~\cite{ho1997,overstreet2022,chiao2023,tobar2024}.

The literature on gravitational effects in quantum systems has generally focused on atomic systems.
In terms of condensed matter and extended systems, a few works on superconductors exist~\cite{dewitt1966,kiefer2005,ummarino2020}.

We  previously explored the effects of gravitation on solid-state quantum information platforms in Ref.~\cite{balatsky2025}, where the analysis was performed on transmon type qubits. Although our analysis was performed considering transmons, the universal nature of gravity means that the results are generically applicable to any qubit type whose computational basis is defined by an energy level splitting. In contrast to earlier works studying gravitational effects on quantum systems, this work studied a solid state system rather than atomic or photonic systems.

In this paper we continue this thread of gravitational effects in solid state systems by analyzing extended systems, specifically those which exhibit topological phases. In particular we consider the Su-Schrieffer-Heeger (SSH) model of a topological insulator \cite{su1979,heeger1988,shortcourse,marino}
and the
Kitaev model of a topological superconductor \cite{kitaev2001}.
In the transmon case the effect of gravitation is a redshift of the energy levels on an individual qubit by gravitational potential in qubit basis. While entangled ensembles of many qubits can be constructed, all experiencing different redshifts, there is no interaction between them.
In the case presented here, not only do the on-site energy levels of the tight-binding lattice become redshifted, but the dynamical tunneling amplitudes between sites is also altered by the gravitational field. This then represents a new category of effects compared to the single transmon case. In a broader context these questions can be viewed as study in the effects of gravity on delocalized and entangled quantum systems.

We note that tight-binding models in curved spacetime have previously been conducted with alternative approximations
\cite{morice2022,rajbongshi2025,marra2025}, however this prior work studied hopping parameters with an inhomogenious modulation mimicking a curved spacetime background rather than modeling an actual gravitational field as in our case here.

The structure of the paper is as follows: In Section~\ref{sec:tb} we derive the effect of gravitation on the tight-binding parameters of our models.
Our main results are presented in Sections~\ref{sec:ssh} and \ref{sec:kitaev}, detailing the effects of gravitation on these topological models, first from analysis of their tight-binding models followed by an analysis of their effective continuum field theories.
We examine the topology of these models in further detail in Section~\ref{sec:topology} by analyzing their symmetries and compute their real-space winding number.
We conclude in Section~\ref{sec:conclusion}. In the Appendices we give further details on the derivations underlying the calculations presented in the main text.

Our key findings are described by Fig.~\ref{fig:dwschematic} and Table~\ref{tab:truth}, which illustrate the gravitationally induced topological phase transition in the Kitaev model and the robustness of the topology of the Majorana states compared to the SSH zero modes.

We use the metric signature $(-,+,+,+)$ and the notation $x^\mu = (x^0,x^j) = (ct,\vec{x})$ where Greek indices span spacetime coordinates and Latin indices span spatial coordinates.

\section{Tight Binding Models\label{sec:tb}}

Condensed matter systems are often described in terms of a tight-binding lattice model. Here we derive the effect of the gravitational potential on the tight-binding parameters.

The spacetime background we use is
\begin{equation}
    \d s^2 = -\qty(1 + \frac{2\Phi(x)}{c^2}) c^2 \d t^2 + \d r^2 + r^2 \d \Omega^2
\end{equation}
where $\d \Omega^2$ is the spherical area element. As we are analyzing systems with only one spatial dimension, we will only consider the time and radial components and use the effective $1+1d$
metric for Newtonian limit
\begin{align}
    \boldsymbol{g}(x)
    &=
    \begin{pmatrix} -1 - \frac{2 \Phi(x)}{c^2}
    & 0 \\ 0 & 1
    \end{pmatrix}
    ,&
    \Phi(x) &= -\frac{\mathrm{G} M}{x_0 + x}
\label{eq:newtonianmetric}
\end{align}
We treat the system in a general-covariant manner by formulating our system in the language of Dirac constraint reparameterization invariance
\cite{dirac,kiefer}.

We start by deriving the non-relativistic limit of the motion of a particle from the action
\begin{align}
	S[x]
	&=	\int \d\tau \left[ -mc\sqrt{ -g_{\mu\nu}(x) \dot{x}^\mu \dot{x}^\nu } - V(\vec{x}) \right]
\end{align}
where the overdot represents derivative with respect to the proper time $\tau$.
This is the action for a relativistic point particle with potential $V(\vec{x})$ in a gravitational background, indicated by the use of the nonconstant spacetime metric in the inner product.
In the limit of low velocity there is no special relativistic dialation, so that $\dot{x}^0 \approx c / \sqrt{-g_{00}(\vec{x})}$.

The canonical Hamiltonian constrant is
\begin{equation}
    c p_0 - \sqrt{-g_{00}(\vec{x})} \sqrt{\vec{p}^{\,2} c^2 + m^2 c^4} - \sqrt{-g_{00}(\vec{x})} V(\vec{x}) \approx 0
\end{equation}
The derivation is outlined in Appendix~\ref{sec:hamiltonian}. 
Under quantization, $p_0 \to -\i\hslash \partial_{x^0}$.
This leads to a Schr\"odinger equation of the form $\left( - \i\hslash \partial_t - H \right) \psi = 0$ 
with Hamiltonian
\begin{equation}
    H = \sqrt{-g_{00}(\vec{x})} \left[ m c^2 + \frac{\vec{p}^{\,2}}{2 m} + V(\vec{x}) + \mathcal{O}(\vec{p}^{\,4}) \right]
\end{equation}
Under quantization we have the Hamiltonian
\begin{equation}
\begin{aligned}
\hat{H}	= Z(x) \left[ mc^2 -\frac{\hslash^2}{2 m} \triangle + V_{\text{lattice}}(r,R) + U(r) \right]
\end{aligned}
\label{eq:redshiftedhamiltonian}
\end{equation}
where we have an overall redshift factor $Z(x) \vcentcolon= \sqrt{1 + \frac{2\Phi(x)}{c^2}}$.
Using the full Schwarzschild metric would involve the Hamiltonian Eq.~\eqref{eq:redshiftedhamiltonian} containing the curved space Laplacian operator in the kinetic $p^2/2m$ term. For simplicity in calculating the tight-binding amplitudes we will not consider this term, but we include it in our analysis of the field theories in Sections~\ref{sec:sshfield} and~\ref{sec:kitaevfield}.


From a many-body Hamiltonian, tight-binding parameters are derived in terms of overlap integrals of the Wannier functions $\phi(r)$ as
\cite{mahan}
\begin{widetext}
\begin{equation}
    \varepsilon_{j}
    =
    \int \dd{r}
    \phi^*(r-R_j)
    \qty[ mc^2 -\frac{\hslash^2}{2 m} \triangle + V_{\text{lattice}}(r,R) ]
    \phi(r-R_j)
\label{eq:wannierenergy}
\end{equation}
and
\begin{equation}
    t_{i,j}
    =
    \int \dd{r}
    \phi^*(r-R_i)
    \qty[ mc^2 -\frac{\hslash^2}{2 m} \triangle + V_{\text{lattice}}(r,R) ]
    \phi(r-R_j) .
\end{equation}
Note that we have retained the rest mass energy $mc^2$ term in the Hamiltonian. This is a scalar, meaning that it drops out of the hopping amplitude as Wannier orbitals are orthogonal and can be considered to be a uniform shift on the on-site energies.

In presence of gravity, as shown above a non-relativistic Hamiltonian becomes redshifted, $H \to \sqrt{-g_{00}(x)} H$. This in turn then redshifts the tight-binding amplitudes, which are derived from the overlap integrals of the Wannier wave functions.

The effect on tight-binding parameters
\begin{align}
    \varepsilon_{j}
    &=
    \int \dd{r}
    \phi^*(r-R_j)
    \sqrt{-g_{00}(r)}
    \qty[ -\frac{\hslash^2}{2 m} \triangle + V_{\text{lattice}}(r,R) ]
    \phi(r-R_j)
\intertext{The gravitational factor can be pulled out of the integral as we can assume $\Phi(r) \approx \Phi(R_n)$ over overlap interval}
    &\approx
    \sqrt{1 + \frac{2\Phi(R_j)}{c^2}}
    \int \dd{r}
    \phi^*(r-R_j)
    \qty[ -\frac{\hslash^2}{2 m} \triangle + V_{\text{lattice}}(r,R) ]
    \phi(r-R_j)
    \\
    &\approx
    \sqrt{ 1 + \frac{2\Phi(R_j)}{c^2} }\,  \varepsilon_j
\end{align}
A similar approximation can be taken for the hopping integral,
\begin{align}
    t_{i,j}'
    &=
    \int \dd{r}
    \phi^*(r-R_i)
    \sqrt{-g_{00}(r)} \left[ -\frac{\hslash^2}{2 m} \triangle + V_{\text{lattice}}(r,R) \right]
    \phi(r-R_j)
    \intertext{where we assume that $\Phi(r)$ can be approximated by its value at the midpoint between the two lattice sites, $r = \frac{R_i+R_j}{2}$}
    &\approx
    \sqrt{1 + \frac{2\Phi(\frac{R_i+R_j}{2})}{c^2}}
    \int \dd{r}
    \phi^*(r-R_i)
    \left[ -\frac{\hslash^2}{2 m} \triangle + V_{\text{lattice}}(r,R) \right]
    \phi(r-R_j)
    \\
    &\approx
    \sqrt{ 1 + \frac{2\Phi(\frac{R_i+R_j}{2})}{c^2} }\, t_{i,j}
\end{align}
Here we see that the $mc^2$ term becomes irrelevant as $\int \d r \phi^*(r-R_i) m c^2 \phi(r-R_j) = 0$ for $i\neq j$ due to the orthogonality of the Wannier functions.
A non-local interaction term can be treated in the same way as the kinetic term
\begin{align}
    U_{m,n}
    &=
    \int \d r \int \d r' \phi^*(r-R_m) \phi(r-R_m)
    U(r,r')
    \phi^*(r'-R_n) \phi(r'-R_n)
\intertext{where similarly for the hopping, the gravitational potential is approximated with its midpoint value}
    &\approx
    \sqrt{ 1 + \frac{2\Phi(\frac{R_m+R_n}{2})}{c^2} }\, U_{m,n}
    .
    \label{eq:Uredshift}
\end{align}


\end{widetext}

As the redshift factor from the metric is a function of the gravitational potential, it is necessary to determine a reference with respect to which the redshift is measured.
\begin{figure}[h]
\centering
\includegraphics[width=\linewidth]{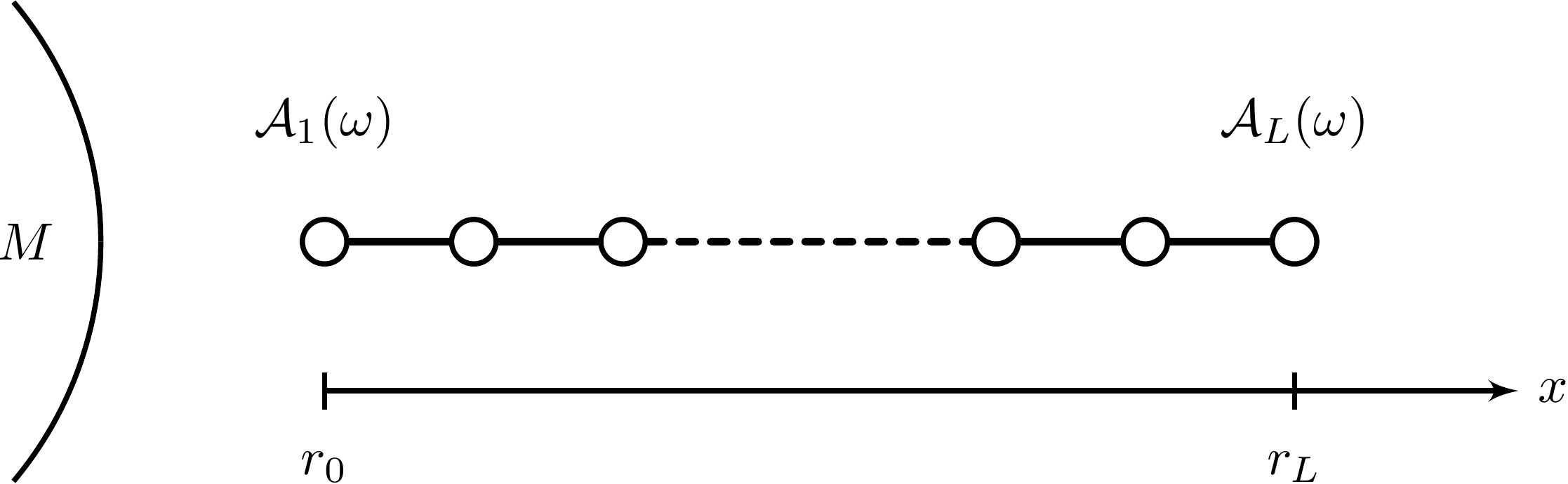}
\caption{Diagram for the system geometry studied. The SSH and Kitaev models are $1d$ chains immersed radially in the gravitational field of a source $M$. One end of the chain is situated at the coordinate $r_0 > R_s$ near the source mass $M$. The terminus of the far end of the chain lies at the coordinate $r_L = r_0 + \ell L$ where $\ell$ is the lattice constant and $L$ is the number of sites in the tight-binding chain. The main quantities of interest are the local spectral functions on the two ends of the chain, $\mathcal{A}_1(\omega)$ and $\mathcal{A}_L(\omega)$. The redshift for a given site is calculated with respect to the gravitational potential at coordinate $r_0$. \label{fig:diagram}}
\end{figure}
The modulation of the Hamiltonian is then taken to be the relative redshift
\begin{align}
    \frac{\sqrt{-g_{00}(x)}}{\sqrt{-g_{00}(0)}}
    &=
    \frac{\sqrt{1 - \frac{2\Phi(x)}{c^2}}}{\sqrt{1 - \frac{2\Phi(0)}{c^2}}}
    .
\end{align}
For paratemeterizing the gravitational effect, we will write the redshift in terms of the Schwarzschild radius $R_S = 2 \mathrm{G} M / c^2$ and declare the distance $x$ with respect to a reference coordinate $r_0$,
\begin{equation}
\frac{\sqrt{-g_{00}(x)}}{\sqrt{-g_{00}(0)}}
    =        \frac{\sqrt{1 - \frac{R_s}{r_0+x}}}{\sqrt{1 - \frac{R_s}{r_0}}}
    .
\end{equation}
In terms of the lattice model parameters derived above, the onsite energies are shifted at lattice site $j$ by $x = j \ell$ and the nearest-neighbor amplitudes are shifted at position $x = (j+\tfrac12) \ell$.

For experiments near the surface of Earth, the parameters are $R_s = \SI{0.008870}{\meter}$ and $r_0 = \SI{6.371E6}{\meter}$. The typical lattice spacing for a solid state system is $\ell \sim \SI{1}{\angstrom}$.
The relative redshift on a realistic solid state system of $L \sim \SI{1}{\micro\meter}$ is approximately $\SI{E-16}{}$. Typical energy scales in solid-state systems is on the order of a few millielectronvolts, meaning that deviations due to the gravitational redshift would be on the order of $\SI{1e-19}{\electronvolt}$, which is far below the accuracy of present experimental measurements.

A more detailed calculation would involve keeping the gravitational potential within the overlap integrals as well as including non-trivial metric terms in the spatial inner product in the kinetic energy term. In the context of the tight-binding models we leave this more detailed analysis for future study, however we will consider these terms in our analysis of the corresponding field theories.

One of the main quantities we wish to obtain is the local spectral function, which we obtain from the retarded Green function.
We use the notation \cite{zubarev1960} $\Green{\op{c}{i}(t')}{\opd{c}{j}(t)} = G_{i,j}(t-t') = -\i \theta(t-t')\left\langle \qty{ \op{c}{}(t') , \opd{c}{}(t) }\right\rangle$ for the retarded Green function as a function of the two times $t$ and $t'$. In the energy representation it is notated as $\Green{\op{c}{i}}{\opd{c}{j}}_{z}$ where $z = \omega + \i \eta$ and $\omega \in \mathbbm{R}$ and $\eta \in \mathbbm{R}^+$.

For our analysis in this paper we will consider the geometry illustrated in Fig.~\ref{fig:diagram} where the $1d$ models will be situated radially from the source mass. For a chain of $L$ sites, site 1 of the chain will be positioned at the coordinate $r_0$ (as measured from the origin) and the opposite end of the chain will lie at site $r_L = r_0 + \ell L$ where $\ell$ is the lattice constant. The source mass is modeled as a point mass at the origin and $r_0$ is always taken to lie outside the event horizon, $r_0 > R_s$.

\section{SSH Model\label{sec:ssh}}

The Su-Schrieffer-Heeger (SSH)model~\cite{su1979,heeger1988,hasan2010,shortcourse} is a paradigmatic model of a symmetry protected topological insulator. The protecting symmetry is that of chiral, or subspace, symmetry. It is conventionally defined as a tight-binding lattice model, but can also be constructed in the form of a continuum field theory. We analyze the effects of including a gravitational background in both of these formulations in the following.

\subsection{Lattice model}

The standard form of the SSH Hamiltonian is given by
\begin{equation}
	\hat{H}_{\textsc{ssh}}
	=
	\sum_{j=1}^{L} \left[ \varepsilon\, \opd{c}{j} \op{c}{j} + \left( \tensor{t}{_A} \opd{c}{2j-1} \op{c}{2j} + \tensor{t}{_B} \opd{c}{2j} \op{c}{2j+1} + \textsc{h.}\text{c.} \right) \right]
    .
\end{equation}
which consists of spinless fermions on a $1d$ lattice with alternating nearest-neighbor hopping amplitudes $t_A$ and $t_B$ and a constant on-site energy $\varepsilon$. The topological phase of the SSH model is given when $t_A < t_B$ with the trivial case being for $t_A \geq t_B$. The primary characteristic of the SSH model is its topological phase is the local spectrum on the boundary exhibiting a gapped band structure with a localized spectral pole at $\varepsilon$, as shown in Fig.~\ref{fig:sshfree}. In the trivial phase the boundary spectrum is gapped and the bulk spectrum is gapped on all sites in both phases.

The SSH Hamiltonian can be taken as a specific implementation of the more general form
\begin{equation}
	\hat{H} = \sum_{j=1}^{L} \left[ \tensor*{\varepsilon}{_j} \opd{c}{j} \op{c}{j} + \tensor*{t}{_{j,j+1}} \left( \opd{c}{j+1} \op{c}{j} + \opd{c}{j} \op{c}{j+1} \right) \right]
\label{eq:basickineticham}
\end{equation}
where the condition for the topological configuration can be expressed as $t_1 < t_2$ or $t_{\text{odd}} < t_{\text{even}}$.
For the standard SSH model, the tunneling amplitudes are given by
\begin{equation}
    t_{j,j+1} = t_0 + (-1)^{j} \delta t
\end{equation}
As described in the previous section, the tunneling amplitudes are redshifted by
\begin{equation}
    t_{j,j+1} = \frac{\sqrt{1 - \frac{R_s}{r_0+\qty(j+\frac12)\ell}}}{\sqrt{1 - \frac{R_s}{r_0}}} \qty[ t_0 + (-1)^{j} \delta t ]
    .
\end{equation}

The Green function equation of motion \cite{zubarev1960} for the Hamiltonian Eq.~\eqref{eq:basickineticham} is\footnote{The equation of motion can be obtained by taking the time derivative of the definition of the retarded Green function, $G_{i,j}(t-t') = -\i \theta(t-t')\left\langle \qty{ \op{c}{}(t') , \opd{c}{}(t) }\right\rangle$, employing the Heisenberg equation of motion for the operator $\opd{c}{}$, and then taking the Laplace transform of the equation to the energy domain. For further details see Ref.~\cite{zubarev1960}.}
\begin{equation}
    ( z - \varepsilon_j ) {G}_{i,j}(z) = \delta_{i,j} + t_{j-1,j} {G}_{i,j-1}(z) + t_{j,j+1} {G}_{i,j+1}(z)
\end{equation}
which can numerically be iterated for the boundary site in the form of a continued fraction
\begin{equation}
	G_{1,1}(z)
	=	\cfrac{1}{z - \varepsilon_{1} - \cfrac{t_{1,2}^2}{z - \varepsilon_{2} - \cfrac{t_{2,3}^2}{{\cfrac{\ddots}{z - \varepsilon_{L-1} - \cfrac{t_{L-1,L}^2}{z - \varepsilon_{L}}}}}}} .
\label{eq:sshcontfrac}
\end{equation}
From this Green function we can obtain the local spectral density of states $\mathcal{A}_j(\omega) = -\frac1\pi\imaginary G_{j,j}(\omega + \i 0^+)$.

\begin{figure}[htp!]
\subfloat[]{
\includegraphics[scale=1]{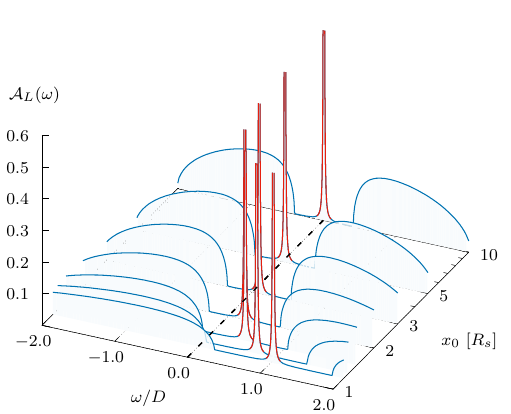}
\label{fig:sshL}
}
\\
\subfloat[]{
\includegraphics[scale=1]{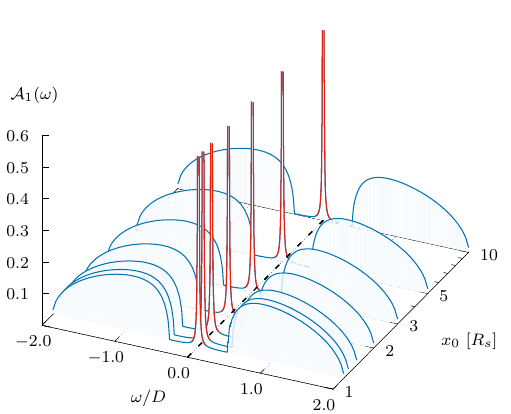}
\label{fig:ssh0}
}
\caption{\protect\subref{fig:sshL} Spectrum on site $L$, $\mathcal{A}_{L}(\omega)$, redshifted with respect to site $1$ of an SSH chain with $\SI{E5}{}$ unit cells in presence of gravity with a source with Schwarzschild radius of $10^5 \ell$ for site $1$ positioned at $x_0 = 1.2 R_s ,\, 1.3 R_s ,\, 1.5 R_s ,\, 2 R_s ,\, 3 R_s ,\, 5 R_s ,\, 10 R_s$. The base band gap is $2\delta t = 0.3D$ in units of the band width $D = 2 t_0$. The SSH gap remains intact even as the shift in energy becomes on the order of relevant energy scales, such as the band width and band gap at site $1$. The topological state follows the observed redshift. The bold dashed line marks $\omega = 0$ where the pole of the gravity-free SSH model lies. The topological spectral pole is highlighted in red. \protect\subref{fig:ssh0} Same sequence of spectral plots for site 1.
\label{fig:ssh}}
\end{figure}

The effect of gravitation on the SSH spectrum is shown in Fig.~\ref{fig:ssh}.
The localized topological state follows the shifted Fermi level and the band width becomes extended.
The SSH band width $D$ is given by $t_A+t_B = 2 t_0$. For the case where the tunneling amplitudes are modified by the gravitational potential, the elevated edge site band width is given by $t_{N-1,N} + t_{N-2,N-1}$. The standard SSH band gap $\Delta$ is given by $t_B - t_A = 2 \delta t$.

Ordinarily, the SSH model is constructed with $\varepsilon_j = 0$ on all sites in order to establish chiral symmetry. However, the zero energy level in a condensed matter system is a matter of convention. The true energy of an electron in a solid is a positive energy value. In order to accurately capture the energy redshift of gravity, we therefore impose an effective positive nonzero Fermi level in the model.
The Fermi level becomes redshifted due to gravity, $\varepsilon_F \to \varepsilon_F \sqrt{1 - \frac{R_s}{r_0+j\ell}}$. If the original Fermil level is normalized to $\varepsilon_F = 0$, then there is no effect of gravitation and the topological state remains at zero. As the real Fermi level is a finite energy, calculations are performed for finite $\varepsilon_F$ and then translated back such that the reference Fermi level lies at 0,
\begin{equation}
    \varepsilon_j = \varepsilon_F \frac{\sqrt{1 - \frac{R_s}{r_0+j\ell}}}{\sqrt{1 - \frac{R_s}{r_0}}} - \varepsilon_F.
\end{equation}
As we will discuss in Section~\ref{sec:topology}, the chiral symmetry can be taken to be with respect to this non-zero $\varepsilon_F$.

In the presence of gravity, the qualitative features of the SSH spectrum remain: two bands separated by a hard gap with a localized state at the midpoint of the gap.
Position of topological spectral pole follows that of the redshifted Fermi level, $\omega^{(g)}_p = \varepsilon^{(g)}_F = \varepsilon_F \frac{\sqrt{1 - \frac{R_s}{r_0+L\ell}}}{\sqrt{1 - \frac{R_s}{r_0}}} - \varepsilon_F$
The exact position of the
outer band edges marking the band width $D$ lie at $\omega^{(g)}_{D} = \varepsilon^{(g)}_F \pm \frac{\sqrt{1 - \frac{R_s}{r_0+\qty(L-\sfrac12)\ell}}}{\sqrt{1 - \frac{R_s}{r_0}}} \omega^{(0)}_{D}$,
and the exact position of the 
inner gap edges are at $\omega^{(g)}_{\Delta} = \varepsilon^{(g)}_F \pm \frac{\sqrt{1 - \frac{R_s}{r_0+\qty(L-\sfrac12)\ell}}}{\sqrt{1 - \frac{R_s}{r_0}}} \omega^{(0)}_{\Delta}$,
where $\omega^{(0)}_{D}={2t_0}$ and ${\omega^{(0)}_{\Delta}}={2 \delta t}$ are the baseline outer and inner band edges respectively.
In our analysis we keep the length of the chain fixed, and change the location of the end of the chain, $x_0$. A diagram of the spectral function at site $L$ redshifted with respect to site~1 positioned at $x_0$ is shown in Fig.~\ref{fig:sshL}.

\subsection{Continuum field theory\label{sec:sshfield}}

We now perform the analysis of the gravitational effects on the SSH model in terms of its continuum field theory~\cite{takayama1980,marino}. The field theory is obtained by taking the long wavelength approximation of the lattice theory and obtaining a Dirac equation. We then put this Dirac equation in curved spacetime to implement the coupling to gravity.

We start from the tight-binding Hamiltonian Eq.~\eqref{eq:basickineticham} and write $j=2m$ for A-sites and $j=2m+1$ for B-sites ($m=1,\ldots,N/2$). We express the alternating hopping terms as
\begin{align}
    t_{2m,2m+1} &= t_0-\delta t =t_1,
    &
    t_{2m+1,2m+2} &= t_0+\delta t=t_2.
\end{align}
We will also allow for a constant on-site energy (in flat space) 
\begin{equation}
    \varepsilon_j = \varepsilon_0.
\end{equation}
Introducing the two-component spinor
\begin{equation}
    \op{\psi}{k} = \begin{pmatrix} \op{c}{A,k} \\ \op{c}{B,k} \end{pmatrix}
\end{equation}
where
\begin{equation}
\begin{aligned}
    \op{c}{A,k} &= \frac{1}{\sqrt{N/2}} \sum_m \e^{-\i km\ell} \op{c}{2m} ,
    \\
    \op{c}{B,k} &= \frac{1}{\sqrt{N/2}} \sum_m \e^{-\i km\ell} \op{c}{2m+1},
\end{aligned}
\end{equation}
the Hamiltonian becomes
\begin{equation}
    \mathcal{H}_{\text{SSH}}(k) = \varepsilon_0 \sigma_0 + [2 t_0 \cos (k\ell)] \sigma_x + [ 2\delta t \sin(k\ell)] \sigma_y. 
\end{equation}
Taking $k = \pi/(2\ell)+q$ with $|q\ell|\ll1$ and expanding to first order, we have 
\begin{equation}
    \mathcal{H}_{\text{SSH}} \approx \varepsilon_0\sigma_0- v_F q \sigma_x + m_0 \sigma_y.
\end{equation}
Here we have defined $v_F \equiv 2 t_0 \ell$ and $m_0 \equiv 2 \delta t$.
Now we take $q \rightarrow - \i \partial_x$ and we additionally rotate the Hamiltonian by the unitary 
\begin{equation}
    U = \exp \left(- \i \frac{\pi}{4}\sigma_x \right) = \frac{1}{\sqrt{2}} (\sigma_0-\i \sigma_x)
\end{equation}
to put the Hamiltonian in the conventional form of
\begin{equation}
    \mathcal{H} = \varepsilon_0 \sigma_0- \i v_F \sigma_x \partial_x + m_0 \sigma_z
\end{equation}
with the mass term proportional to $\sigma_z$.

To write the action, we choose the Clifford algebra basis to be
\begin{align}
    \gamma^t &= \sigma_z ,& \gamma^x &= \i \sigma_y
\label{eq:sshcliffflat}
\end{align}
such that $\{\gamma^\mu , \gamma^\nu\} = 2 \eta^{\mu\nu}$. We can then write the action for the field theory in flat spacetime as\footnote{We will write $t$ for the $x^0$ component with the understanding that $x^0$ is really $ct$, \textit{i.e.} $\partial_t \equiv \frac{\partial}{c\partial t}$.}
\begin{equation}
    S = \frac{1}{2} \int \d t \d x\, \overline{\Psi}\qty[ \i \gamma^t \partial_t + \i v_F \gamma^x\partial_x - m_0 - \varepsilon_0 \gamma^t]\Psi. 
\end{equation}

From here, we now put the theory on a curved spacetime background of the form 
\begin{equation}
    \d s^2 = - Z^2(x) c^2 \d t^2 + Z^{-2}(x) \d x^2 
\end{equation}
where for condensed notation we use $Z(x) = \sqrt{1 + \frac{R_s}{x}}$ and $x$ labels the radial coordinate of the exterior Schwarzschild metric. In terms of the orthogonal coframe 1-form basis $\vartheta^a$ \cite{hehl1990,parkertoms}, the line element is expressed as $g_{\mu\nu} \d x^\mu \otimes \d x^\nu = \eta_{ab} \vartheta^a \otimes \vartheta^b$. The coordinate zweibein frame components $\tensor{e}{^a_\mu}$ are given by $\vartheta^a = \tensor{e}{^a_\mu} \d x^\mu$. Greek indices label spacetime coordinates taking values in $\{t,x\}$ and latin indices label the orthogonal frame basis taking values in $\{0,1\}$.
As the metric is diagonal we can pick the zweibein to be
\begin{align}
    \tensor{e}{^0_t} &= Z(x), & \tensor{e}{^1_x} &= Z^{-1}(x), 
\label{eq:zweibein}
\end{align}
with all other zweibein $=0$.

The connection 1-forms $\tensor{\omega}{^a_b}$ can be determined from the torsion-free Cartan structure equation
\begin{equation}
    \d \vartheta^a + \tensor{\omega}{^a_b} \wedge \vartheta^b = 0.
\end{equation}
Explicitly, in our case we therefore have 
\begin{align}
    \tensor{\omega}{^0_1_t} &= Z(x) Z'(x)
    &
    \tensor{\omega}{^0_1_x}&=0.
\label{eq:connections}
\end{align}
From these we derive the components of the spin connection by
\begin{equation}
    \tensor{\Omega}{_\mu} = \frac{1}{4} \tensor{\omega}{^a^b_\mu} [\gamma_a, \gamma_b] ,
\end{equation}
where $\tensor{\omega}{^a_c} = \eta_{bc} \tensor{\omega}{^{ab}_\mu} \d x^\mu$.
The Clifford algebra basis in the orthogonal frame inherit the form from the flat spacetime basis Eq.~\eqref{eq:sshcliffflat}, $\gamma^0 = \sigma_z$ and $\gamma^1 = \i \sigma_y$.
For the components of the spin connection we have 
\begin{align}
        \Omega_t &= \frac{1}{2} \tensor{\omega}{^0^1_t} \sigma_x = \frac{1}{2} Z Z' \sigma_x , 
 & \Omega_x &= 0.
\label{eq:spinconnections}
\end{align}
For the Lagrangian we now transition from flat spacetime to curved spacetime by making the replacement $\slashed{\partial}\rightarrow \slashed{D} = \tensor{e}{_a^\mu}\gamma^a \qty( \partial_{\mu}+\Omega_{\mu} )$, yielding
\begin{equation}
    \begin{aligned}
    \mathcal{L}
    &= \overline{\Psi}[\i \tensor{e}{_a^\mu} \gamma^a (\partial_{\mu}+\Omega_{\mu})-m_0-\varepsilon_0 \gamma^0 \tensor{e}{_0^t}] \Psi
    \\
    &= \overline{\Psi} \left[ \frac{\i}{Z} \sigma_z \partial_t - v_F Z \sigma_y \partial_x - \frac{Z'}{2} \sigma_y - m_0 - \frac{\varepsilon_0}{Z} \sigma_z \right] \Psi . 
    \end{aligned}
\end{equation}
This Lagrangian then leads to the Hamiltonian
\begin{equation}
    \mathcal{H}(x) = \varepsilon_0 \sigma_0 - \i v_F Z^2 \sigma_x \partial_x - \i \frac{1}{2} Z Z' \sigma_x + Z m_0 \sigma_z .
\end{equation}
The retarded Green function $G^R(\omega;x,x')$ satisfies 
\begin{equation}
    [\omega + \i 0^+ - \mathcal{H}(x)] G^R (\omega;x,x') = \delta(x-x') 1. 
\end{equation}
For the zero-mode solution, we look for a bound state at energy $\omega=E$ obeying 
\begin{equation}
    [E - \mathcal{H}(x)] \tilde{\Psi}_E(x)=0.
\end{equation}
From the Hamiltonian we can obtain the zero mode solution
\begin{equation}
    \frac{E-\varepsilon_0}{Z(x)} \tilde{\Psi}_E(x)  = [ \i v_F Z(x) \sigma_x + \i \frac{1}{2} Z'  \sigma_x \partial_x - m_0 \sigma_z ] \tilde{\Psi}_E(x).
\label{eq:sshzerostate}
\end{equation}
In this basis the chiral symmetry operator takes the form $\Gamma = \sigma_y$, and we have
\begin{equation}
    \{ \Gamma, \sigma_x \} = \{ \Gamma , \sigma_z \} = 0.
\end{equation}
The right hand side of Eq.~\eqref{eq:sshzerostate} satisfies chiral symmetry. In order for the left hand side to satisfy it as well, it must vanish. This pins the energy to 
\begin{equation}
    E = \varepsilon_0. 
\end{equation}
Note that the factor of $Z(x)$ is irrelevant here, because it's an overall multiplicative factor. What this reflects is that a \textit{local} observer does not see the redshift. An observer at infinity would see energy shifted by 
\begin{equation}
    E = \frac{\varepsilon_0}{Z(x_0)}. 
\end{equation}
This manifests as the usual redshift, which reflects the dialation observed in the analysis of the tight-binding model above.
The time dilation factor appears here as an inverse as the field theory is analyzed using an observer at spatial infinity, where as the tight-binding model was analyzed using an observer at the end of the chain closest to the gravitational source.

In order to provide a direct comparison with the analysis of the tight-binding model, we can consider the case where we take the approximation such that $\tensor{e}{^1_x} = 0$.
Ignoring these spatial components of the metric, the zero energy bound state solution reads as
\begin{equation}
    \frac{E-\varepsilon_0}{Z(x)} \tilde{\Psi}_E(x)
    =
    [ \i v_F \sigma_x \partial_x - m_0 \sigma_z ] \tilde{\Psi}_E(x) ,
\end{equation}
which possess the same qualitative behavior for the redshift of the bound state as Eq.~\eqref{eq:sshzerostate}. We can therefore conclude that the analysis of the tight-binding model with the harder Newtonian approximation accurately describes the topological feature in the presence of gravity.

\section{Kitaev Wire\label{sec:kitaev}}

Just as the SSH model is a paradigmatic model of a $1d$ topological insulator, the Kitaev wire is similarly a paradigmatic model of a symmetry protected $1d$ topological superconductor~\cite{kitaev2001,hasan2010}. As with the SSH model, the Kitaev model has a protecting symmetry which is particle-hole symmetry.
%
%
\subsection{Lattice model}
The Kitaev model is a $1d$ $p$-wave superconductor with Hamiltonian
\begin{equation}
\begin{aligned}
	\hat{H} = \sum_{j=1}^{L}  \Big[ &- t_{j,j+1} \left( \opd{c}{j} \op{c}{j+1} + \opd{c}{j+1} \op{c}{j} \right) - \mu_{j} \left( \opd{c}{j} \op{c}{j} - \frac12 \right) \\&+ \Delta_{j,j+1} \op{c}{j} \op{c}{j+1} + \Delta^*_{j,j+1} \opd{c}{j+1} \opd{c}{j} \Big]
\end{aligned}
\end{equation}
in terms of spinless fermions where $\Delta_{i,j}$ is the superconducting pairing function obeying the $p$-wave symmetry $\Delta_{j,i} = - \Delta_{i,j}$.
This system can be rewritten in terms of Majorana degrees of freedom
to obtain the form
\begin{equation}
\begin{aligned}
    H
    =
    \frac\i2 \sum_{j=1}^{L}
    \Big[ 
        \mu_{j} \gamma_{j,1} \gamma_{j,2}
		&+ ( t_{j,j+1} - \Delta_{j,j+1} ) \gamma_{j,1} \gamma_{j+1,2} \\&+ ( - t_{j,j+1} - \Delta_{j,j+1} ) \gamma_{j,2} \gamma_{j+1,1} 
    \Big]
\label{eq:kitaevhamiltonian}
\end{aligned}
\end{equation}
where the relation between the fermionic and Majorana degrees of freedom is
\begin{align}
	\gamma_{j,1} &\vcentcolon= \op{c}{j} + \opd{c}{j} \,,
	&
	\gamma_{j,2} &\vcentcolon= \frac{\op{c}{j} - \opd{c}{j}}{\i}
	&
	j &= 1,\ldots,L
    .
\end{align}
As with the SSH model, the Kitaev model exhibits a topological phase which is characterized by a gapped local spectral function on the boundary with a zero energy spectral pole, the Majorana zero mode (MZM), shown in Fig.~\ref{fig:kitaevfree}. Unlike the SSH model, the Kitaev topological pole is pinned to exactly zero energy, rather than the fermionic on-site energy $\mu$.

The tight-binding parameters of the Hamiltonian Eq.~\eqref{eq:kitaevhamiltonian} transform by a relative redshift
\begin{subequations}
\begin{align}
    \mu_{j} &= \frac{\sqrt{1 - \frac{R_s}{r_0+j\ell}}}{\sqrt{1 - \frac{R_s}{r_0}}}
    \, ( \mu_{0} + \varepsilon_F ) - \varepsilon_F
    \\
    t_{j,j+1} &= \frac{\sqrt{1 - \frac{R_s}{r_0+(j+\frac12)\ell}}}{\sqrt{1 - \frac{R_s}{r_0}}}
    \, t_{0}
    \\
    \Delta_{j,j+1} &= \frac{\sqrt{1 - \frac{R_s}{r_0+(j+\frac12)\ell}}}{\sqrt{1 - \frac{R_s}{r_0}}}
\, \Delta_{0}
    \label{eq:Deltaredshift}
\end{align}
\end{subequations}
For direct comparison with the analysis of the SSH model, a ficticious chemical potential $\varepsilon_F$ is added to the on site energies and then subtracted after the redshift.

\begin{figure}[htp!]
\subfloat[]{\includegraphics[scale=1]{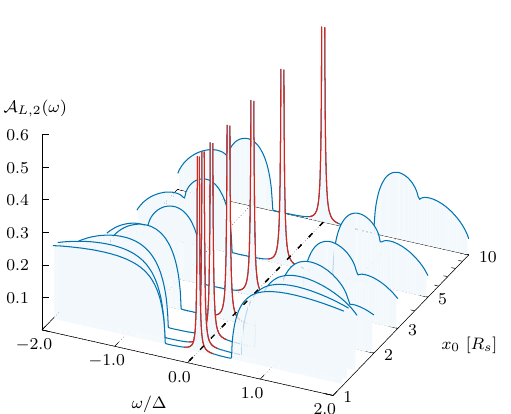}\label{fig:kitaevL}}
\\
\subfloat[]{
\includegraphics[scale=1]{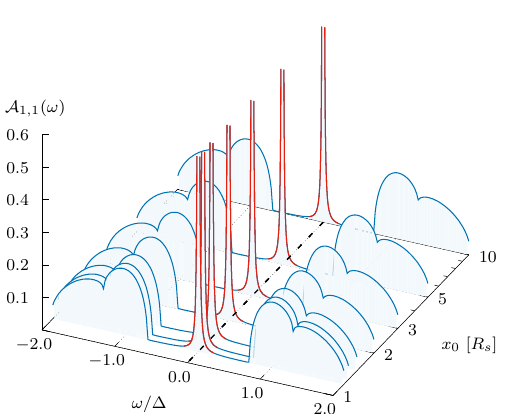}
\label{fig:kitaev0}
}
\caption{
\protect\subref{fig:kitaevL} Spectrum of $\gamma_2$ on site~$L$ redshifted with respect to site~$1$ of a Kitaev wire with $\SI{E5}{}$ Majorana unit cells in presence of gravity for site $1$ positioned at $x_0 = 1.2 R_s ,\, 1.3 R_s ,\, 1.5 R_s ,\, 2 R_s ,\, 3 R_s ,\, 5 R_s ,\, 10 R_s$, with $R_s = 10^{5}\ell$. The system parameters are $\mu_0/\Delta_0 = 0.3$, $t_0/\Delta_0 = 0.5$, $\varepsilon_F/\Delta_0 = 1$, $R_s = 10^{5}\ell$. The topological state, highlighted in red, is observed to remain at $\omega = 0$ for various values of $x_0$, in contrast to the SSH model where the topological state tracks with the observed redshift. \protect\subref{fig:kitaev0}~Same sequence of spectral plots for $\gamma_1$ on site~1.
\label{fig:kitaev}}
\end{figure}

As for the SSH, we obtain the local spectral function by means of the retarded Green function, which we calculate from its equation of motion. 
The Green function for the Majorana degrees of freedom, can be expressed by
\begin{equation}
	\mathcal{G}_{i,j}(z)
	=
	\begin{pmatrix}
		\Green{ \gamma_{i,1} }{ \gamma_{j,1} }_z	&	\Green{ \gamma_{i,2} }{ \gamma_{j,1} }_z
		\\
		\Green{ \gamma_{i,1} }{ \gamma_{j,2} }_z	&	\Green{ \gamma_{i,2} }{ \gamma_{j,2} }_z
	\end{pmatrix}
\end{equation}
The Green function equation of motion is given by
\begin{equation}
\begin{multlined}
    \left[ z\mathbbm{1} - \mathcal{E}_{j} \right] \mathcal{G}_{i,j}(z) =
    \\	
    \shoveright{2\delta_{i,j}\mathbbm{1} + \mathcal{T}^-_{j-1,j} \mathcal{G}_{i,j-1}(z) + \mathcal{T}^+_{j,j+1} \mathcal{G}_{i,j+1}(z)}
\end{multlined}
\end{equation}
where
\begin{subequations}
\begin{align}
	\mathcal{E}_{j}		&=	\i \begin{pmatrix} 0 & - \mu_j \\ \mu_j & 0 \end{pmatrix}
	\\
	\mathcal{T}^-_{j-1,j}	&=	-\i \begin{pmatrix} 0 & \Delta_{j-1,j} + t_{j-1,j} \\ \Delta_{j-1,j} - t_{j-1,j} & 0 \end{pmatrix}
	\\
	\mathcal{T}^+_{j,j+1}	&=	\i \begin{pmatrix} 0 & \Delta_{j,j+1} - t_{j,j+1} \\ \Delta_{j,j+1} + t_{j,j+1} & 0 \end{pmatrix}
\end{align}
\end{subequations}
Similarly to the SSH model, the solution for the Green function $\mathcal{G}_{1,1}(z)$ for a given set of parameters $( \{\mu_{j}\} , \{\Delta_{j}\} , \{t_{j}\} )$ is obtained recursively. Further details of this calculation are presented in Appendix~\ref{sec:green}.
The Majorana density of states can then be calculated from the Green function
\begin{equation}
	\mathcal{A}_{1,1}(\omega) = -\frac1\pi \imaginary \Green{ \gamma_{1,1} }{ \gamma_{1,1} }_{\omega^+} .
\end{equation}
Similarly to the SSH model, the topological state manifests as a spectral pole at zero energy. The spectrum of the Majorana at site $L$ redshifted with respect to site~1 is given by
\begin{equation}
	\mathcal{A}_{L,2}(\omega) = -\frac1\pi \imaginary \Green{ \gamma_{L,2} }{ \gamma_{L,2} }_{\omega^+} .
\end{equation}
This spectrum for varying positions of $x_0$ is depicted in Fig.~\ref{fig:kitaev}.
The boundary Majorana states are robust in presence of weak gravity. The zero energy feature in spectrum persists. 
The gap edge of the Majorana spectrum however does shift and lies at $2t_j-\mu_j$.
%

In contrast to the SSH model, the redshift in energies due to gravitation affects the band gap, allowing gravitation to facilitate a topological phase transition by closing the gap.
Also in contrast to the SSH model, the topological state remains fixed to zero energy, despite the fact that the actual Fermi level is redshifted.

\subsection{Gravitationally induced topological phase transition\label{sec:phasetransition}}

\begin{figure}[htp!]
\centering
\subfloat[]{
\includegraphics[width=\linewidth]{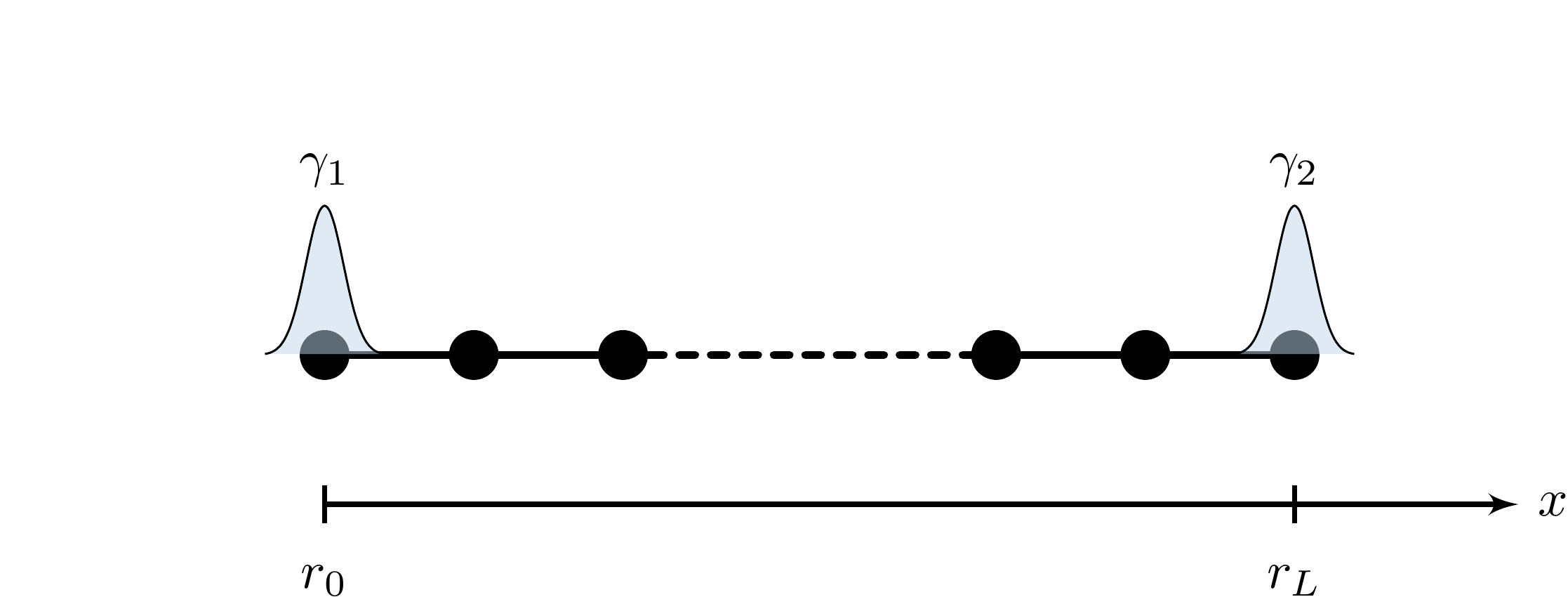}
\label{fig:dwschematic0}
}
\\
\subfloat[]{
\includegraphics[width=\linewidth]{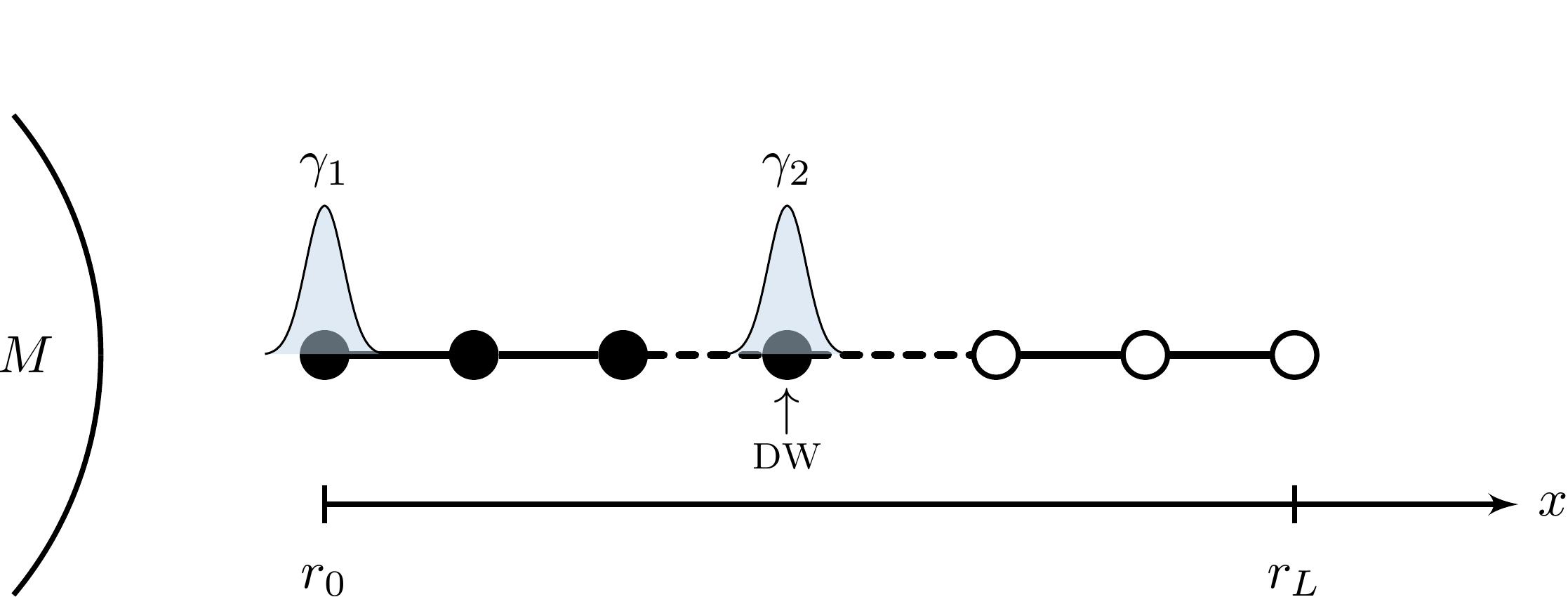}
\label{fig:dwschematic1}
}
\caption{Gravitationally induced topological phase transition. Schematic demonstrating the localization of the Majorana zero modes (MZM) $\gamma_1$, $\gamma_2$ on the tight-binding chain. Filled sites represent the topological phase. \protect\subref{fig:dwschematic0} In the flat spacetime case the MZM are localized on the end points of the chain and the entire system is in the topological phase. \protect\subref{fig:dwschematic1} Under certain parameterization, in the presence of gravity the second Majorana $\gamma_2$ becomes localized on a domain wall within the bulk of the chain. The domain wall partitions the system into a part which is in the topological phase and a part in the trivial phase, illustrated by the open sites. \label{fig:dwschematic}}
\end{figure}

The phases of the Kitaev wire are parameterized by the condition
\begin{equation}
    \begin{cases}
        \mu < 2 t   &   \text{Topological}
        \\
        \mu \geq 2 t    &   \text{Trivial}
    \end{cases}
\end{equation}
Extrapolating this to spatially dependent parameters in the presence of gravity, the phase transition occurs at position $X$ when
\begin{equation}
    \frac{\sqrt{1 - \frac{R_s}{r_0+X}}}{\sqrt{1 - \frac{R_s}{r_0}}} (\mu_0 + \varepsilon_F) - \varepsilon_F
    =
    2 \frac{\sqrt{1 - \frac{R_s}{r_0+X+\frac12}}}{\sqrt{1 - \frac{R_s}{r_0}}}\, t_0
\label{eq:domainwallcondition}
\end{equation}
such that the ratio $\mu(x) / 2t(x)$ changes sign for $x<X$ and $x>X$.
Under an appropriate parameterization of $r_0, R_s, \mu_0, \varepsilon_F, t_0$, we find that this phase transition can occur at an $X$ which lies within the bulk of the wire, leading to the creation of a domain wall~\cite{shortcourse,sen2020,wong2022,wongthesis}.
When a domain wall forms the system becomes split into two regions, one which remains topological and one which becomes trivial. Accordingly, the Majorana state becomes localized around the domain wall at the boundary of the topological region rather than the terminus of the chain. Qualitatively the boundary of the chain can itself be interpreted as a domain wall between a topological region and a trivial region, where the trivial region is the vacuum.
This case can be seen schematically in Fig.~\ref{fig:dwschematic}, where the Majorana wavefunction shifts from being localized on the end of the chain, to a bulk site where the domain wall sits.

\begin{figure}[htp!]
    \includegraphics[scale=1]{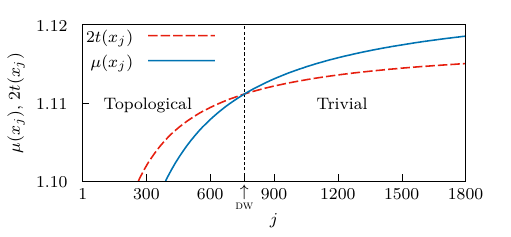}
    \caption{Magnitude of the redshifted Kitaev wire parameters $\mu(x)$ and $t(x)$ along their position on the wire. System parameters are $\mu_0/\Delta_0 = 0.9$, $t_0/\Delta_0 = 0.5$, $\varepsilon_F / \Delta_0 = 1$, $R_s = 10\ell$, $x_0=5R_s$. Initially the system lies in the topological phase with $\mu(x_j) < 2 t(x_j)$, however their relative magnitude swaps at the domain wall transitioning the remainder of the system into the trivial phase. The domain wall occurs at $j_{\textsc{dw}} = 760$ as indicated by the arrow and vertical dashed line. \label{fig:DWconditionplot}}
\end{figure}

As a concrete example, we consider the parameterization $\mu_0/\Delta_0 = 0.9$, $t_0/\Delta_0 = 0.5$, $\varepsilon_F / \Delta_0 = 1$, $R_s = 10\ell$, and $r_0 = 5 R_s$. This parameterization puts the system close to the topological phase transition, with $\mu$ and $2t$ close in value. 
These conditions lead to the formation of a domain wall as shown by Fig.~\ref{fig:DWconditionplot} where at the head of the chain $\mu(x) < 2t(x)$ putting the system in the topological phase, but at a point in the bulk of the system this ratio reverses and $\mu(x) < 2t(x)$ in the remainder of the system. For this parameterization the domain wall occurs at site $j=760$.

\begin{figure}[htp!]
\subfloat[\label{fig:dwlengthlin}\hspace{2.7em}\phantom{.}]{
\includegraphics[scale=1]{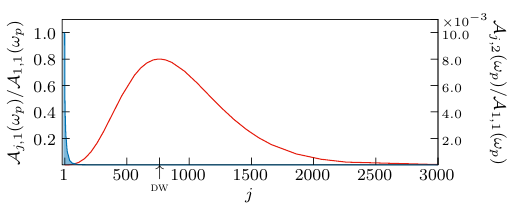}\label{fig:dwlengthlin}}
\\
\subfloat[\label{fig:dwlengthlog}\hspace{2.7em}\phantom{.}]{
\includegraphics[scale=1]{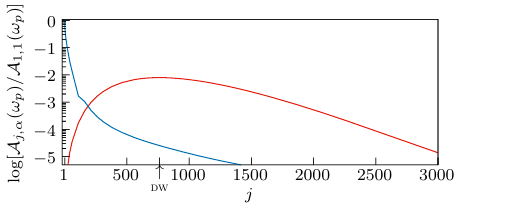}\label{fig:dwlengthlog}}
\caption{Magnitude of the Majorana spectral pole for $x_0 = 5R_s$, yielding a domain wall at $j_{\textsc{dw}}=760$ according to Table~\ref{tab:splitting} and indicated by the arrows on the plots. The $\gamma_1$ (blue) is localized on the boundary at site 1 and $\gamma_2$ (red) is localized on the bulk domain wall site 760. System parameters are $\mu_0/\Delta_0 = 0.9$, $t_0/\Delta_0 = 0.5$, $\varepsilon_F / \Delta_0 = 1$, $R_s = 10\ell$. Magnitude of the spectral peak is normalized to the magnitude of $\gamma_1$ on site~1. Spectra are plotted \protect\subref{fig:dwlengthlin} on a linear scale, where the scale for the $\gamma_2$ spectrum is on the left axis, and \protect\subref{fig:dwlengthlog} on a log scale, where now both spectra are plotted using the same axis. \label{fig:dwlength}}
\end{figure}

\begin{figure}[htp!]
\centering
\subfloat[]{
\includegraphics[scale=1]{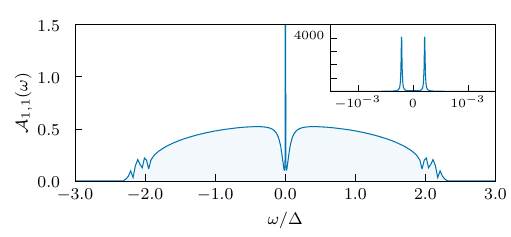}
\label{fig:DW01}
}
\\
\subfloat[]{
\includegraphics[scale=1]{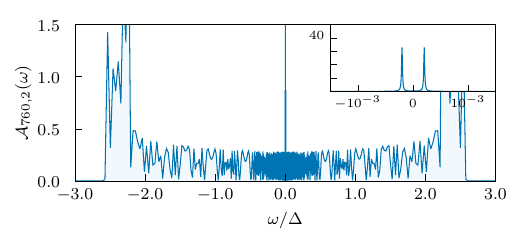}
\label{fig:DW7602}
}
\caption{Spectral functions of \protect\subref{fig:DW01} $\gamma_1$ on the boundary site and \protect\subref{fig:DW7602} $\gamma_2$ on the domain wall site.
\textcolor{red}{(Insets)} show the splitting of the poles due to their hybridization. Note the difference in magnitude of the poles. The domain wall site in the bulk approaches the structure of the bulk spectrum in the trivial partition, \textit{cf.} Fig.~\ref{fig:DWbulk}. System parameters are $\mu_0/\Delta_0 = 0.9$, $t_0/\Delta_0 = 0.5$, $\varepsilon_F / \Delta_0 = 1$, $R_s = 10\ell$, $x_0 = 5 R_s$. \label{fig:DWtop}}
\end{figure}

\begin{figure}[htp!]
\centering
\subfloat[]{
\includegraphics[scale=1]{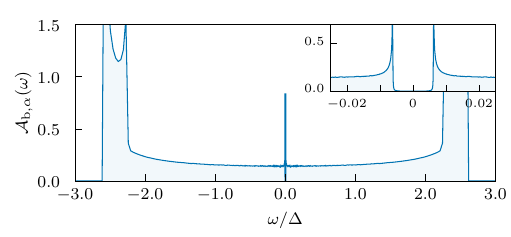}
\label{fig:DWbulk}
}
\\
\subfloat[]{
\includegraphics[scale=1]{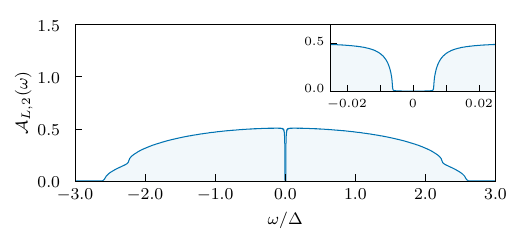}
\label{fig:DWL2}
}
\caption{Spectral functions for the Kitaev wire in a gravitational background with $x_0 = 5R_s$. \protect\subref{fig:DWbulk} Bulk cell in the trivial side of the chain $\mathrm{b}=49630$ which is half-way between the domain wall and the far boundary $L = \num{2E5}$; $\alpha=1,2$ is the Majorana index. The spectrum is the same for both Majoranas. \protect\subref{fig:DWL2} Spectral function for $\gamma_2$ on the $L$ cell. Given the presence of a spectral peak for $\gamma_1$ on cell 1, a complementary spectral peak would ordinarily be expected here, but the peak has moved to the domain wall at $j=760$, leaving this spectrum topologically trivial and gapped. \textcolor{red}{(Insets)} are zoomed into the region near zero energy showing the gap and absence of a topological pole. \label{fig:DWtriv}}
\end{figure}

The local spectra for this system is shown in Figs.~\ref{fig:DWtop} and~\ref{fig:DWtriv}. Fig.~\ref{fig:DWtop} shows the topological spectral pole of $\gamma_1$ on the boundary (Fig.~\ref{fig:DW01}) and of $\gamma_2$ on the domain wall (Fig.~\ref{fig:DW7602}). The topological states appear as a pair of poles as their wavefunctions overlap, causing them to hybridize.
The wave functions of the MZM in the presence of the domain wall are plotted in Fig.~\ref{fig:dwlength}.
Fig.~\ref{fig:DWtriv} shows the spectrum of sites beyond the domain wall in the trivial partition of the chain. Unlike in the usual domain wall-free case, the spectrum for $\gamma_2$ on site $L$ (Fig.~\ref{fig:DWL2}) is trivially gapped, owing to the fact that this MZM now lies on the domain wall site.

The creation of the domain wall results in overlapping wavefunctions of the Majorana states, leading to a splitting in their energy levels. The energy level splitting is a function of the distance within the chain at which the domain wall forms. Numerical data for this is presented in Table~\ref{tab:splitting}.

There exists a subtle dependency on the ficticious Fermi level $\varepsilon_F$. In order for the gravitationally induced phase transition to occur, the added $\varepsilon_F$ must satisfy
\begin{equation}
    \varepsilon_F > \frac{1}{1 - \sqrt{\frac{1}{1 - \frac{R_s}{r_0}}}} (2 t_{0} - \mu_{0}) .
\end{equation}

\begin{table}[htp!]
\caption{Relationship between the proximity of the end of a Kitaev chain to a gravitational source $x_0$ with the location of the domain wall within the chain $j_{\textsc{dw}}$ as calculated from Eq.~\eqref{eq:domainwallcondition}, and the location of the topological spectral poles as obtained from $\mathcal{A}(\omega)$.
System parameters are $\mu_0/\Delta_0 = 0.9$, $t_0/\Delta_0 = 0.5$, $\varepsilon_F / \Delta_0 = 1$, $R_s = 10\ell$.
The poles in the spectrum $\mathcal{A}(\omega)$ appear at $\omega = \pm \omega_p$ as the spectrum is symmetric. The relationship between the location of the domain wall and the energy splitting of the topological state approximately obeys $|\omega_p| \sim 10^{-\ln(j_{\textsc{dw}})/2}$.
\label{tab:splitting}}
\begin{tabular}{|c|c|c|}
\hline
\phantom{.}\quad $r_0 [R_s]$ \quad\phantom{.} & \phantom{.}\quad $j_{\textsc{dw}}$ \quad\phantom{.} & \phantom{.}\quad $\omega_p / \Delta_0$ \quad\phantom{.}
\\\hline
5.0 & 760 & 0.000209 
\\\hline
4.5 & 206 & 0.003186 
\\\hline
4.0 & 95 & 0.010586 
\\\hline
3.5 & 50 & 0.023206 
\\\hline
3.0 & 27 & 0.043452 
\\\hline
2.5 & 14 & 0.076836 
\\\hline
2.0 & 6 & 0.138333 
\\\hline
\end{tabular}
\end{table}


\subsection{Continuum field theory\label{sec:kitaevfield}}

As with our analysis of the SSH model, we procede to construct the field theory in curved spacetime for the Kitaev chain in order to incorporate the nontrivial spatial components of the metric.
We start from the lattice Hamiltonian Eq.~\eqref{eq:kitaevhamiltonian}.
An important distinction between the two models is that the zero-modes of the SSH model sit at the Fermi energy, whereas MZMs sit at exactly zero energy in any frame, because BdG Hamiltonians are written relative to $\mu$ by construction. We move into the BdG description by introducing the Nambu spinor 
\begin{equation}
    \op{\psi}{k} =\begin{pmatrix} \op{c}{k} \\ \opd{c}{-k} \end{pmatrix}
\end{equation}
such that the Hamiltonian takes the BdG form
\begin{equation}
    \op{H}{} = \sum_k \opd{\psi}{k} \mathcal{H}_{\text{BdG}}(k) \op{\psi}{k},
\end{equation}
with
\begin{equation}
    \mathcal{H}_{\text{BdG}}(k)= [-2 t \cos (k\ell) - \mu]\tau_z + 2 \Delta \sin(k\ell) \tau_x.
\end{equation}
Here $\tau_i$ are Pauli matrices in the internal Nambu space and $\ell$ is the lattice constant. 
If we expand for $k\ell \ll 1$ and expand around the Fermi points via $k = \pm k_F + q$ with $|q\ell|\ll 1$ and keep only first-order in $q$, this leads to the Hamiltonian taking the form
\begin{equation}
    \mathcal{H}(x) = -\i v_F \tau_z \partial_x + m \tau_y
\end{equation}
where $m \equiv 2 \Delta$ and $v_F = 2 t \ell \sin (k_F\ell)$. 
Our Clifford algebra basis takes the form
\begin{equation}
    \gamma^0 = \tau_x , \qquad \gamma^1 = \i \tau_y, \qquad \{ \gamma^{\mu},\gamma^{\nu}\} = -2\eta^{\mu\nu} .
\end{equation}
In this prescription, we reproduce the flat-space Hamiltonian via 
\begin{equation}
    S = \frac{1}{2} \int \d t \d x \overline{\Psi} \qty(\i \gamma^0 \partial_t + \i v_F \gamma^1 \partial_x - m ) \Psi, \qquad m \equiv \mu. 
\end{equation}
Here the real spinor $\Psi^{\dagger} = \Psi^{\mathsf{T}} C \gamma^0$ obeys the Majorana condition $\Psi = C \overline{\Psi}^{\mathsf{T}}$. 

We now move to curved space with 
\begin{equation}
    \d s^2 = - Z^2(x) c^2 \d t^2 + Z^2(x) \d x^2
\end{equation}
as before, with the zweibein, connection, and spin connection components as in Section~\ref{sec:sshfield}, Eqs.~(\ref{eq:zweibein},\ref{eq:connections},\ref{eq:spinconnections}).
From this, we have the curved spacetime action 
\begin{equation}
    S = \frac{1}{2} \int \d t \d x\, \overline{\Psi} \left[ \frac{\i}{Z} \sigma_x \partial_t - v_F Z \sigma_y \partial_x + \frac{Z'}{2} \sigma_y - m \right] \Psi. 
\end{equation}
By multiplying the equation of motion by $Z \sigma_x$, this curved space action leads to the Hamiltonian
\begin{equation}
    \mathcal{H} = - \i v_F Z \sigma_z \partial_x - \frac{\i \partial_x Z}{2} \sigma_z - m Z \sigma_x .
\end{equation}


To analyze the topological state, we take the kink profile for the mass
\begin{equation}
    m(x) = \mu(x) = \mu_0 \sign(x), \qquad |\mu_0|<2t. 
\end{equation}
We can define a weighted spinor $\chi(x)$ by 
\begin{equation}
    \Psi(x) = Z^{-1/2}(x) \chi(x)
\end{equation}
and the spatial derivative becomes
\begin{equation}
    \begin{split}
        \partial_x \Psi & = (\partial_x Z^{-1/2})\chi + Z^{-1/2} \partial_x \chi \\
        & = - \frac{ \partial_x Z}{2 Z^{3/2}} \chi + \frac{1}{\sqrt{Z}} \partial_x \chi.
    \end{split}
\end{equation}
Thus, the term $Z \partial_x + (\partial_x)Z/2$ becomes
\begin{equation}
    \begin{split}
        Z \partial_x \Psi + \frac{1}{2} \partial_x Z \Psi & = Z \left( - \frac{ \partial_x Z}{2 Z^{3/2}} \chi + \frac{1}{\sqrt{Z}} \partial_x \chi \right) + \frac{ \partial_x Z}{2} \frac{\chi}{\sqrt{Z}} \\
        & = - \frac{ \partial_x Z}{2 \sqrt{Z}} \chi + \sqrt{Z} \partial_x \chi + \frac{\partial_x Z}{2 \sqrt{Z}} \chi \\
        & = \sqrt{Z} \partial_x \chi. 
    \end{split}
\end{equation}
The stationary state equation $\mathcal{H} \Psi = E \Psi$ therefore becomes
\begin{equation}
    -[\i v_F \sigma_z \partial_x - m(x) \sigma_x] \chi = \frac{E}{Z(x)} \chi. 
\end{equation}
Again, the overall factor of $Z(x)$ is irrelevant.  

The important difference here is that there is no equivalent of an $\varepsilon_0$ term---so there cannot be a redshift in the zero mode. The zero-energy modes are zero-energy even for a distant observer.

\section{Topology\label{sec:topology}}

In the general sense, a topological feature implies a feature which is independent of the metric, or in other words, independent of gravity. For example, action of Chern-Simons theory, the prototypical example of a topological field theory with action
\begin{equation}
    S_{\text{CS}}[A] = \int_M A \wedge \d A = \int_M \dd[3]{x} \varepsilon^{ijk} A_i \partial_j A_k \,,
\end{equation}
is metric independent with $\delta S_{\text{CS}}/\delta g_{\mu\nu}$ vanishing identically.
Formally speaking neither the field theories of the SSH model or the Kitaev model are topological field theories in this sense, as their actions explicitly depend on the metric by way of the vielbeine. However, we would nevertheless expect a topological feature to be independent of the metric. In the cases of the SSH and Kitaev models, this means the localized zero energy state. As shown in the previous sections, the behavior of the topological state in the two models differ in their response to a gravitational background. The energy level of the topological spectral pole of the SSH model becomes redshifted. In contrast, the spectral pole of the Kitaev wire remains at the same energy level. We can therefore say that the SSH model is not truly topological, but the Kitaev model is. In this section we outline some quantitative reasoning for this.

For symmetry protected topological matter, such as the SSH and Kitaev models, the notion of topology and the system's symmetries are correlated. We therefore analyze these two systems's symmetries as well as their topological invariant winding number.

\subsection{Symmetry}

Both the SSH model and the Kitaev wire are topological systems whose topology is protected by chiral symmetry \cite{hasan2010,shortcourse}. 
In the sublattice basis the chiral symmetry operator is written as
\begin{equation}
    \mathcal{S} = \begin{pmatrix} 1 & 0 \\ 0 & -1 \end{pmatrix} \otimes \mathbbm{1}_N
\end{equation}
where $N$ is the number of unit cells of the system.
For SSH and Kitaev this basis takes the form of $\begin{pmatrix} A \\ B \end{pmatrix}$, where in SSH $A$ and $B$ correspond to the $A$ and $B$ lattice sites, and in the Kitaev wire they correspond to the $\gamma_{1}$ and $\gamma_{2}$ of each fermion site.
A Hamiltonian obeys chiral symmetry if
\begin{equation}
    \mathcal{S}^{-1} H \mathcal{S} = -H .
\label{eq:chiralsymmetry}
\end{equation}
For the SSH model, chiral symmetry is only satisfied if the diagonal terms of the Hamiltonian vanish, \textit{i.e.} if $\varepsilon_i = 0$ $\forall i$. For a constant shift in the on site energies, such as by uniform doping, an effective chiral symmetric Hamiltonian $\tilde{H}$ can be obtained by $\tilde{H} = H - \varepsilon \mathbbm{1}$. However, for inhomogeneous on site energies, this uniform shift to obtain a chiral symmetric Hamiltonian cannot be performed. This is the case here with gravitation where the local on site energies become redshifted according to their position. The SSH Hamiltonian $H$ in the presence of gravity does not satisfy Eq.~\eqref{eq:chiralsymmetry}, nor does any shifted Hamiltonian $\tilde{H}$, and is therefore not chiral symmetric.
The implications of this symmetry breaking are that the localized midgap state is no longer required to be fixed to zero energy. The symmetry protecting the topological state is now broken.

The situation regarding chiral symmetry for the Kitaev model is different. The Kitaev model in the Majorana basis does preserve the chiral symmetry, even in the presence of gravity where the local energies are redshifted. In the Majorana basis, there are no terms along the diagonal of the Hamiltonian. In matrix notation, the Kitaev Hamiltonian Eq.~\eqref{eq:kitaevhamiltonian} reads as
\begin{equation}
    \mathcal{H}_{\textsc{k}} = 
    \begin{pmatrix}
        0 & \i\mu_1 & 0 & -\frac\i2 T^-_1 &
        \\
        -\i\mu_1 & 0 & -\frac\i2 T^+_1 & 0 & \ddots
        \\
        0 & \frac\i2 T^+_1 & 0 & \i\mu_2 & \ddots
        \\
        \frac\i2 T^-_1 & 0 & -\i\mu_2 & 0 & \ddots
        \\
        & \ddots & \ddots & \ddots & \ddots
    \end{pmatrix}
\end{equation}
where $T^-_j=\Delta_j - t_j$ and $T^+_j=\Delta_j + t_j$.
From this, it is clear that $\mathcal{H}_{\textsc{k}}$ satisfies Eq.~\eqref{eq:chiralsymmetry}, regardless of the site dependent values of the parameters $\mu_j$, $t_j$, and $\Delta_j$.
The Kitaev model therefore retains the symmetry protecting its topological state.

\subsection{Winding Number}
The topology of the $1d$ SSH and Kitaev models is conventionally calculated by means of the Zak phase \cite{zak1989}. However, as this is a momentum space quantity, it cannot be used in the present case due to the spatial inhomogeneity introduced by the gravitational field.
In order to characterize the topology we therefore compute the real space winding number~\cite{mondragonshem2014,jin2024}.

We consider a lattice with unit cells $x = 1 , \ldots, L$, with two sites per cell. The sublattices are labeled $A$ and $B$. The basis kets are therefore $\ket{x,A}$, $\ket{x,B}$. The chiral (sublattice) operator $\Gamma = \sigma_z$ in the $(A,B)$ basis, such that $\Gamma \ket{x,A}=\ket{x,A}$ and $\Gamma \ket{x,B}=-\ket{x,B}$. The projectors onto the two sublattices are 
\begin{align}
    P_A &= \frac{1}{2} (1+\Gamma), & P_B &= \frac{1}{2} (1-\Gamma). 
\end{align}
The position operator on the unit cells is defined as 
\begin{equation}
    \hat{X} = \adjustlimits{\sum^L}_{x=1}{\sum}_{\alpha \in \{A,B\} } x \ketbra{x,\alpha}{x,\alpha}.
\end{equation}
In this state vector representation the SSH Hamiltonian is 
\begin{equation}
    \hat{H} = \sum_{x=1}^L \qty[ t_A \ketbra{x,A}{x,B} + t_B \ketbra{x+1,A}{x,B} + \textsc{h.}\text{c.} ]
\end{equation}
with intracell hopping $t_A$ and intercell hopping $t_B$. Every term flips $A \leftrightarrow B$, so $\{ \Gamma , H \} = 0$ (in other words, all matrix elements that are diagonal in the sublattice basis vanish). The SSH model is in the chiral class, so the AIII real-space winding construction from the papers applies. 

Similarly, the Kitaev Hamiltonian in the Majorana basis in the state vector representation reads as
\begin{equation}
    \hat{H}
    =
    \begin{multlined}[t]
    \frac\i2 \sum_{x=1}^{L} [ \mu \ketbra{x,A}{x,B} - (\Delta-t) \ketbra{x,A}{x+1,B} \\- (\Delta+t) \ketbra{x,B}{x+1,A} + \textsc{h.}\text{c.} ]
    \end{multlined}
\end{equation}
where now $A/B$ label the Majoranas $\gamma_1 / \gamma_2$. This Hamiltonian also posseses chiral symmetry and belongs to class BDI, which has the same topological invariant as class AIII.

Now, let $H = U \mathrm{diag}(E_n) U^{\dagger}$, with $E_n\neq0$. We're treating an infinite chain here; for a finite open chain, exact zero modes are $O(1)$ and will not be relevant. We can define the flat-band operator 
\begin{equation}
    Q \equiv \frac{H}{|H|} = U \diag[\sign(E_n)]U^{\dagger} = P_+ - P_-. 
\end{equation}
Here, $P_{\pm}$ are the spectral projectors above/below zero. Because $\{\Gamma,H\}=0$, we also have $\{\Gamma,Q\}=0$. This means $Q$ is off-diagonal in the $\Gamma$ basis 
\begin{equation}
    Q = \begin{pmatrix} 0 & Q_0 \\ Q_0^{\dagger} & 0 \end{pmatrix} ,
\end{equation}
with
\begin{align}
    Q_0 &\equiv P_A Q ,& P_B Q_0^{-1} &= P_B Q P_A.
\end{align}
The inverse identity is the one used to put the winding in covariant real-space form. The one-dimensional chiral winding in real space is 
\begin{equation}
    \nu = -\Tr_{\mathrm{vol}} \qty{ Q_0^{-1} [X, Q_0] }, 
\end{equation}
where $\mathrm{Tr}_{\mathrm{vol}}$ is the trace per unit length. For a finite chain of length $L$, we use 
\begin{equation}
    \nu_L \equiv -\frac{1}{L} \Tr\qty{ (P_B Q P_A)[X,P_A Q P_B] },
\label{eq:rswn}
\end{equation}
which converges rapidly to the integer $\nu$ away from the critical point. This works in real space because the derivative $\partial_k$ in the usual Bloch formula becomes $-\i [ X, \cdot]$ in real space, and the $A/B$ off-diagonal block $Q_0$ plays the role of the normalized complex map whose phase winds, and the trace per volume replaces $\frac{1}{2\pi} \int \d k$. 

Using Eq.~\eqref{eq:rswn} we can characterize the topology of the SSH and Kitaev models in curved spacetime. Considering a finite size system consisting of 100 unit cells, for both of the systems in the presence of gravity the winding number remains quantized in their respective topological phases. In other words, we have
\begin{align}
    &\text{SSH with }
    t_{1} < t_{2}
    &
    \nu_{\textsc{ssh}} &= 1
    \label{eq:sshwinding}
\intertext{and}
    &\text{Kitaev with }
    \mu_{1} < 2 t_{1}
    &
    \nu_{\textsc{k}} &= 1
    .
    \label{eq:kitaevwinding}
\end{align}
These winding numbers remain constant regardless of the position $x_0$.

A caveat here is that for short chains deep in the gravitational well, \textit{e.g.} $x_0 = 1.2$ and $L\sim100$, the winding number for the Kitaev model is actually $\nu_{\textsc{k}}=0$, even when the parameters suggest the system should be in the topological phase. This is due to finite size effects where the overlap between the boundary Majorana wave functions results in a hybridization which causes a splitting in the zero energy state. The winding number then reads case as zero. In contrast, for the SSH model this hybridization and energy degeneracy splitting does not happen in short systems, as the boundary states do not coincide at the same energy. The winding number is therefore quantized to a finite value, even though the chiral symmetry is broken. For sufficiently short chains of both models the winding number vanishes, even in the absence of gravity, due to the overlap of the gapless boundary mode wave functions.

We summarize this section with Table~\ref{tab:truth}. Unlike in the flat spacetime case, topological invariants and symmetries are no longer correlated in the presence of gravity. While for the Majorana Kitaev wire the protecting symmetry and the winding number are both preserved, for the SSH model only the winding number is preserved in the presence of gravity. This is despite the fact that the Hamiltonian no longer respects the symmetry which protects the topology in flat spacetime.
\begin{table}
\caption{Truth table indicating the preservation of the winding number $\nu$ and satisfying of the symmetry $\mathcal{S}$ for the Kitaev and SSH models with or without a background gravitational field. A $\checkmark$ represents that the winding number is preserved or that the symmetry is satisfied. For the Kitaev model both $\nu$ and $\mathcal{S}$ are preserved in the presence gravity compared to the gravity-free case. For the SSH model only the winding number is preserved. The SSH model in a gravitational background no longer satisfies chiral symmetry. \label{tab:truth}}
\includegraphics[width=0.7\linewidth]{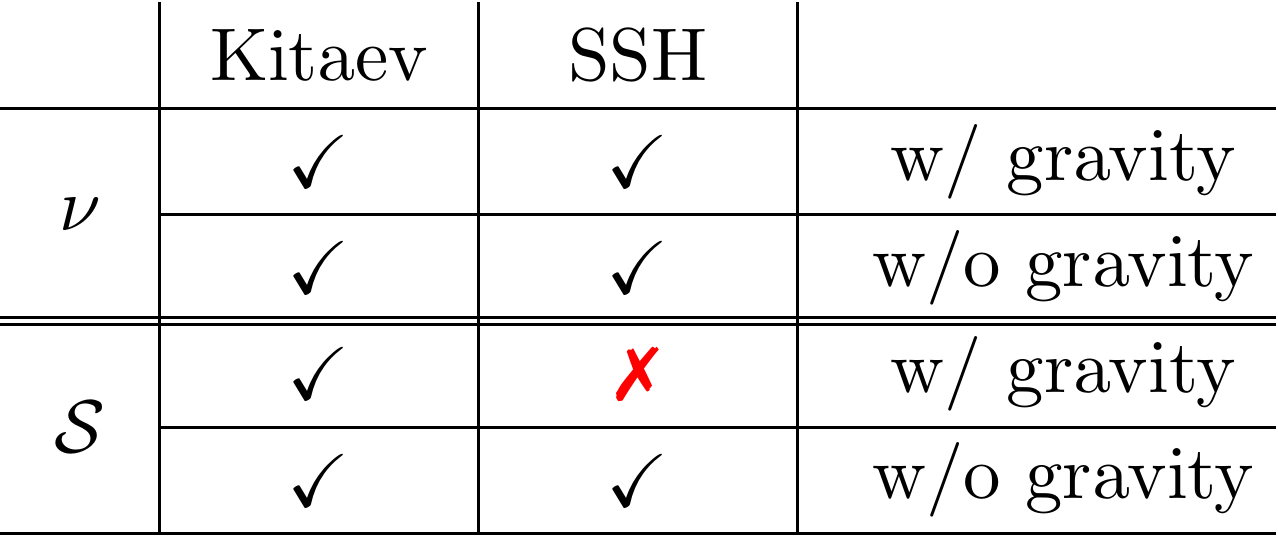}
\end{table}

\section{Conclusion\label{sec:conclusion}}

In this paper we analyzed the effects of a weak gravitational field on the Su-Schrieffer-Heeger (SSH) model and Kitaev wire, paradigmatic 1$d$ models of a topological insulator and a topological superconducter.
The key findings are described by Fig.~\ref{fig:dwschematic} and Table~\ref{tab:truth}.

Described by Table~\ref{tab:truth}, we find that the response of the topology to the gravitational field differs between the two examples.
In the SSH model the energy of the topological state is pinned to the Fermi energy, which becomes redshifted due to gravity. This means that the energy of the topological states of a vertically aligned chain differ by the redshift factor.
In contrast, the topological Majorana states of the Kitaev wire remain at the same energy. The energy of both Majorana states remains the same irrespective of the presence of the gravitational field.

In spite of these differences, the real-space winding number quantifying the topology of the systems remains quantized for both models in their respective topological phases, see Eqs.~\eqref{eq:sshwinding} and~\eqref{eq:kitaevwinding}.

Described by Fig.~\ref{fig:dwschematic} in Section~\ref{sec:phasetransition}, we observe in the Kitaev model a gravitationally induced topological phase transition where the gravitational field is able to drive a phase transition within the the bulk of the system by creating a domain wall.
Such an appearance of a domain wall has impact on the quantum information properties of the Majorana zero modes as qubit states, as it is known that energy level splittings lead to a dephasing channel~\cite{boross2022,peeters2024}. On the other hand, since gravitation does not itself produce any deviations in the energy levels of the MZMs, we do not expect gravitation alone to be a significant dephasing channel in this context.

More broadly, our work demonstrates the appearance of non-trivial effects in condensed matter systems in non-inertial frames, which are similarly described by non-constant spacetime metrics.

\acknowledgements{We thank A.~Bestwick, D.~Pappas, J.~T.~Heath and M.~Zych for useful discussions. This work was supported by the Knut and Alice Wallenberg Foundation Grant No. KAW 2019.0068, the European Research Council under the European Union Seventh Framework Grant No. ERS-2018-SYG 810451 HERO, and the University of Connecticut.}

\appendix
\numberwithin{figure}{section}

\section{Relativistic Hamiltonian\label{sec:hamiltonian}}
In this appendix we review the relation between the incorporation of gravity in Hamiltonians by way of a metric and by that of the addition of a potential.
The action for a relativistic particle is given by the worldline Lagrangian
\begin{align}
    S[x]
    &= - m c \int \d s
    \\
    &= m c \int \sqrt{-\eta_{\mu\nu} \d x^\mu \d x^\nu}
    \\
    &= m c^2 \int \d t \sqrt{-\eta_{\mu\nu} \d \dot{x}^\mu \d \dot{x}^\nu}
    \\
    &= m c^2 \int \d t \sqrt{1 - \qty(\frac{\d\vec{x}}{c\d t})^2}
    .
\end{align}
Performing a series expansion in powers of $c$ yields the Lagrangian
\begin{equation}
    L_0
    =
    m c^2 \sqrt{1 - \qty(\frac{\d\vec{x}}{c\d t})^2}
    =
    m c^2 + \frac{m}{2} \qty(\frac{\d\vec{x}}{\d t})^2 + \mathcal{O}(c^{-2})
\end{equation}
which, modulo the factor of $mc^2$, is clearly the nonrelativistic Lagrangian for a free particle. A potential can be incorporated by $L_0 \to L_0 - V(\vec{x})$, for example the Newtonian gravitational potential $V = \mathrm{G} M m / x$. Now instead of simply adding this potential term, we use the curved spacetime metric $g_{\mu\nu} = \mathrm{diag}\qty[1+2\Phi(x)/c^2,1,1,1]$. This means that the Lagrangian now takes the form of
\begin{equation}
    L = m c^2 \sqrt{1 + \frac{2\Phi(x)}{c^2} - \qty(\frac{\d\vec{x}}{c\d t})^2}
\end{equation}
which to lowest order in $c$ yields
\begin{equation}
    L = m c^2 + \frac{m}{2} \qty(\frac{\d\vec{x}}{\d t})^2 - m \Phi(x)
\end{equation}
which is equivalent to the usual nonrelativistic Lagrangian for a particle in the presence of the gravitational scalar potential $\Phi(x) = - \mathrm{G}M/x$.

In the main text, we use the full $\sqrt{1+\frac{2\Phi(x)}{c^2}}$ term, 
however, in the analysis of the tight-binding models we ignore contributions from the spatial components of the metric, which to order of $1/c^4$ includes the terms
\begin{equation}
    H^{(1)} = \Phi(x) \frac{p^2}{m^2 c^2} - \Phi(x) \frac{p^4}{2 m^3 c^4} - \Phi(x)^2 \frac{p^2}{m^2 c^4}
\end{equation}
beyond the terms which come from the overall redshift factor.

The action for a relativistic particle in curved spacetime is
\begin{align}
	S[x]
	&=	- m c \int_\gamma \d s
	\\
	&=	- m c \int \d\tau \sqrt{ -g_{\mu\nu} \dot{x}^\mu \dot{x}^\nu }
\end{align}
The canonical momentum may be obtained in the usual manner from $p_\mu = {\partial L}/{\partial \dot{x}^\mu}$, yielding
\begin{align}
	p_\mu &=	\frac{m c}{\sqrt{-\dot{x}^2}} g_{\mu\nu} \dot{x}^\nu ,
	&
	\dot{x}^\mu &= \frac{\sqrt{-\dot{x}^2}}{m c} g^{\mu\nu} p_\nu .
\end{align}
In order to perform the Legendre transformation to obtain the canonical Hamiltonian, we first seek an expression for $\sqrt{-\dot{x}^2}$ in terms of the momenta. We can take the $\mu=0$ component of the canonical momentum to write
\begin{align}
    p_0
	&=	\frac{m c}{\sqrt{-\dot{x}^2}} g_{0\nu} \dot{x}^\nu
    =	\frac{m c}{\sqrt{-\dot{x}^2}} g_{00} \dot{x}^0
\label{eq:p0}
\end{align}
where the second equality follows for nonrotating metrics.
The component $p_0$ can also be obtained using the mass shell condition as
\begin{align}
	p_0 &= \pm\sqrt{-g_{00}}\sqrt{\vec{p}^{\,2} + m^2 c^2}
\end{align}
Plugging this expression into Eq.\eqref{eq:p0} yields
\begin{align}
	\sqrt{-\dot{x}^2}
	&=	\pm \frac{m c \sqrt{-g_{00}} \dot{x}^0}{\sqrt{\vec{p}^{\,2} + m^2 c^2}}
\end{align}
The Legendre transformation to obtain the canonical Hamiltonian can now be performed
\begin{align}
	H_C	&=	p_\mu \dot{x}^\mu - L
	\\
	&=	p_\mu \dot{x}^\mu - m c \sqrt{ -g_{\mu\nu} \dot{x}^\mu \dot{x}^\nu }
	\\
	&=	p_0 \dot{x}^0 + p_j \dot{x}^j - m c \sqrt{-\dot{x}^2}
	\\
	&=	p_0 \dot{x}^0 - \frac{\sqrt{-\dot{x}^2}}{m c} p_i g^{ij} p_j - \frac{m^2 c^2 \sqrt{-g_{00}} \dot{x}^0}{\sqrt{\vec{p}^{\,2} c^2 + m^2 c^4}}
	\\
	&=	\dot{x}^0 \left[ p_0 - \frac{\sqrt{-g_{00}} \vec{p}^{\,2}}{\sqrt{\vec{p}^{\,2} c^2 + m^2 c^4}} - \frac{\sqrt{-g_{00}} m^2 c^2}{\sqrt{\vec{p}^{\,2} c^2 + m^2 c^4}} \right]
	\\
	&=	\dot{x}^0 \left[ p_0 - \frac1c \sqrt{-g_{00}} \sqrt{\vec{p}^{\,2} c^2 + m^2 c^4} \right]
\end{align}
The canonical Hamiltonian takes the form of a constraint with $\dot{x}^0$ playing the role of a Lagrange multiplier.
The radical involving the spatial momenta can be expanded in powers of $c$ to obtain a nonrelativistic approximation.
For a Lagrangian involving a potential, $V(x)$, this term may also be included in the Hamiltonian by factoring out a $\dot{x}^0$. 

\section{Redshift of the lattice Kitaev Hamiltonian}
Here we provide some details deriving the redshift of the superconducting pairing amplitude $\Delta_{j,j+1}$ in Eq.~\eqref{eq:Deltaredshift}. This is obtained from constructing a BCS-type mean-field theory~\cite{mahan}.
\begin{widetext}
We start from the Hamiltonian for spin-polarized fermions with nearest-neighbor interactions of
\begin{align}
	\op{H}{}
	&=
		\sum_{n=1}^{L} \left[ \mu_n \left( \opd{c}n \op{c}n - \tfrac12\right) 
		+ t_n \left( \opd{c}{n+1} \op{c}{n} + \opd{c}{n} \op{c}{n+1} \right) 
		+ U_{j,j+1} \opd{c}{j} \opd{c}{j+1} \op{c}{j+1} \op{c}{j} \right]
\end{align}
A Hubbard-Stratonovich transformation on the interaction term results in
\begin{align}
    \tensor{U}{_{j,j+1}} \opd{c}{j} \opd{c}{j+1} \op{c}{j+1} \op{c}{j}
    \to
    \frac{1}{2 U_{j,j+1}} |\Delta_{j,j+1}|^2 - \Delta_{j,j+1} \opd{c}{j} \opd{c}{j+1} - \Delta_{j,j+1} \op{c}{j+1} \op{c}{j}
    \label{eq:hubbardstratonovich}
\end{align}
which leads to the Kitaev model of a $1d$ topological superconductor \cite{kitaev2001}
\begin{equation}
	\hat{H} = \sum_{j=1}^{L}  \left[ - t_{j,j+1} \left( \opd{c}{j} \op{c}{j+1} + \opd{c}{j+1} \op{c}{j} \right) - \mu_{j} \left( \opd{c}{j} \op{c}{j} - \frac12 \right) + \Delta_{j,j+1} \op{c}{j} \op{c}{j+1} + \Delta^*_{j,j+1} \opd{c}{j+1} \opd{c}{j} \right] .
\label{eq:kitaevhamiltonianfermion}
\end{equation}
This system can be rewritten in terms of Majorana degrees of freedom
\begin{align}
	\gamma_{j,1} &\vcentcolon= \op{c}{j} + \opd{c}{j}
	&
	\gamma_{j,2} &\vcentcolon= \frac{\op{c}{j} - \opd{c}{j}}{\i}
	&
	j &= 1,\ldots,L
\end{align}
%
to obtain the form
\begin{align}
    H
    &=
    \frac\i2 \sum_{j=1}^{L}
    \Big[ 
        \mu_{j} \gamma_{j,1} \gamma_{j,2}
		+ ( t_{j,j+1} - \Delta_{j,j+1} ) \gamma_{j,1} \gamma_{j+1,2} + ( - t_{j,j+1} - \Delta_{j,j+1} ) \gamma_{j,2} \gamma_{j+1,1} 
    \Big] .
\end{align}
From Eq.~\eqref{eq:Uredshift} we understand that the interaction amplitude scales as $U_{j,j+1} \to \sqrt{-g_{00}(j+\tfrac12)}\, U_{j,j+1}$.
We can therefore conclude from Eq.~\eqref{eq:hubbardstratonovich} that the superconducting pairing amplitude transforms as
\begin{align}
    \Delta_{j,j+1} &\to \sqrt{-g_{00}(j+\tfrac12)}\, \Delta_{j,j+1}
\end{align}
Similarly to the hopping amplitude, the $p$-wave pairing amplitude is nonlocal between nearest-neighbors. The scaling of the superconducting pairing amplitude then goes with the center of mass coordinate for the gravitational potential.

\section{Kitaev lattice Green functions\label{sec:green}}

Here we present some details on solution for the Kitaev model Green functions from the equation of motion.
\begin{figure}
\subfloat[\label{fig:sshfree}]{\includegraphics[scale=1]{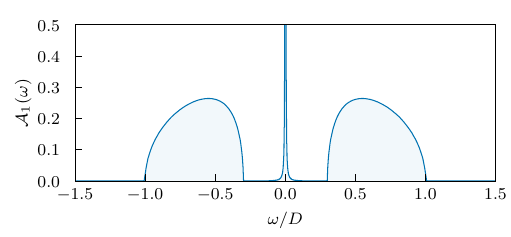}\label{fig:sshfree}}
\subfloat[\label{fig:kitaevfree}]{\includegraphics[scale=1]{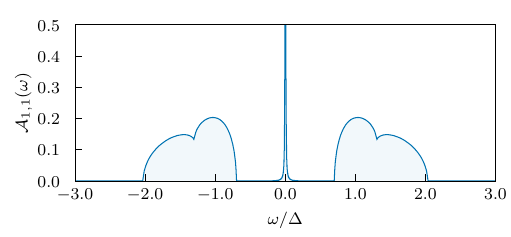}\label{fig:kitaevfree}}
\\
\subfloat[\label{fig:kitaevfree2}]{\includegraphics[scale=1]{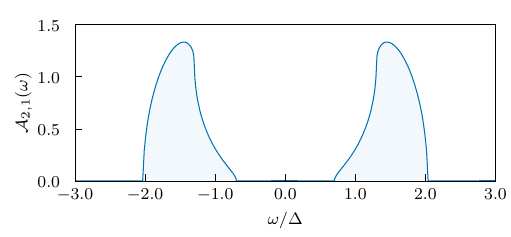}\label{fig:kitaevfree2}}
\subfloat[\label{fig:kitaevfreebulk}]{\includegraphics[scale=1]{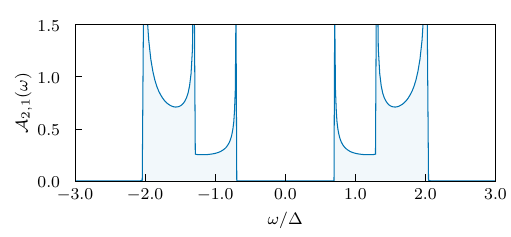}\label{fig:kitaevfreebulk}}
\caption{
Boundary site spectral function in the absence of gravity illustrating the topological phase for \protect\subref{fig:sshfree} the SSH model with $2\delta t = 0.3 D$ and \protect\subref{fig:kitaevfree},\protect\subref{fig:kitaevfree2} the Kitaev wire with $\varepsilon = 0.3 \Delta$ and $t = 0.5 \Delta$. Panel \protect\subref{fig:kitaevfree} shows the spectral function for $\gamma_1$ and panel \protect\subref{fig:kitaevfree2} shows the spectral function for $\gamma_2$. The gap in the $\gamma_2$ spectrum shows that there is only one Majorana localized on the boundary of the chain, $\gamma_1$. For a finite chain the $\gamma_2$ Majorana would be localized on the opposite boundary with the $\gamma_1$ spectrum fully gapped on that site. The difference magnitude of the spectrum of the bands reflects the pole weight of the Majorana zero mode. Spectral functions are obtain from the local Green function which are calculated using a numerical fast recursion algorithm \cite{lopezsancho1984,lopezsancho1985} for a semi-infinite chain. \label{fig:freespec}
}
\end{figure}
As seen by Eq.~\eqref{eq:sshcontfrac} the Green function for a $1d$ Hamiltonian with nearest-neighbor hopping is simple to compute and takes the form of a continued fraction. In contrast, the Kitaev wire does not only contain nearest-neighbor couplings in the Majorana basis, so the Green function does not take the form of a simple continued fraction, but rather a matrix one. For a chain of length $N$, the recursion is initialized using the equation of motion for Green function relating the first unit cell and the last unit cell
\begin{align}
	\left[ z\mathbbm{1} - \mathcal{E}_{N} \right] \mathcal{G}_{1,N}	&=	\mathcal{T}^-_{N-1,N} \mathcal{G}_{1,N-1}
\end{align}
as $\mathcal{G}_{1,N+1}(z) = 0$ due to the boundary conditions. This equation is rearranged into the matrix ratio
\begin{align}
	\mathcal{G}_{1,N} \mathcal{G}_{1,N-1}^{-1}	&=	\left[ z\mathbbm{1} - \mathcal{E}_{N} \right]^{-1} \mathcal{T}^-_{N-1,N}
\end{align}
The equation of motion for the Green function relating the first cell to a general cell $j$ can similarly be rearranged into a ratio
\begin{align}
	\mathcal{G}_{1,j} \mathcal{G}_{1,j-1}^{-1}	&=	\left[ z\mathbbm{1} - \mathcal{E}_{j} - \mathcal{T}^+_{j,j+1} \mathcal{G}_{1,j+1} \mathcal{G}_{1,j}^{-1} \right]^{-1} \mathcal{T}^-_{j-1,j}
\end{align}
The recursion relation is propagated until the ratio $\mathcal{G}_{1,2} \mathcal{G}_{1,1}^{-1}$ is obtained, from which the final Green function $\mathcal{G}_{1,1}$ can be calculated from the equation of motion as
\begin{align}
	\mathcal{G}_{1,1}	&=	2 \left[ z\mathbbm{1} - \mathcal{E}_{1} - \mathcal{T}^+_{1,2} \mathcal{G}_{1,2} \mathcal{G}_{1,1}^{-1} \right]^{-1}
\end{align}
For a bulk site $x$ the equation of motion is
\begin{align}
	\left[ z\mathbbm{1} - \mathcal{E}_{x} \right] \mathcal{G}_{x,x}(z)	&=	2 \mathbbm{1} + \mathcal{T}^-_{x-1,x} \mathcal{G}_{x,x-1}(z) + \mathcal{T}^+_{x,x+1} \mathcal{G}_{x,x+1}(z)
	\\
	\mathcal{G}_{x,x}(z)	&=	2 \left[ z\mathbbm{1} - \mathcal{E}_{x} - \mathcal{T}^-_{x-1,x} \mathcal{G}_{x,x-1} \mathcal{G}_{x,x}^{-1} - \mathcal{T}^+_{x,x+1} \mathcal{G}_{x,x+1} \mathcal{G}_{x,x}^{-1} \right]^{-1}
    .
\end{align}
The solution can be obtained by iterating the Green functions from 1 to $x$ starting from
\begin{align}
	\mathcal{G}_{x,1} \mathcal{G}_{x,2}^{-1}	&=	\left[ z\mathbbm{1} - \mathcal{E}_{1} \right]^{-1} \mathcal{T}^+_{1,2}
\end{align}
and continuing with
\begin{align}
	\mathcal{G}_{x,j} \mathcal{G}_{x,j+1}^{-1}	&=	\left[ z\mathbbm{1} - \mathcal{E}_{j} - \mathcal{T}^-_{j-1,j} \mathcal{G}_{x,j-1} \mathcal{G}_{x,j}^{-1} \right]^{-1} \mathcal{T}^+_{j,j+1}
\end{align}
until $j=x-1$. The remaining term iterates starting from $N$
\begin{align}
	\mathcal{G}_{x,N} \mathcal{G}_{x,N-1}^{-1}	&=	\left[ z\mathbbm{1} - \mathcal{E}_{N} \right]^{-1} \mathcal{T}^-_{N-1,N}
	\\
	\mathcal{G}_{x,j} \mathcal{G}_{x,j-1}^{-1}	&=	\left[ z\mathbbm{1} - \mathcal{E}_{j} - \mathcal{T}^+_{j,j+1} \mathcal{G}_{x,j+1} \mathcal{G}_{x,j}^{-1} \right]^{-1} \mathcal{T}^-_{j-1,j}
\end{align}
and iterating until $j = x+1$.
\end{widetext}

\bibliography{references}

\end{document}